\documentclass[groupedaddress,showkeys]{revtex4-2}
\usepackage{float}
\usepackage{multirow}
\usepackage{amsmath}
\usepackage{amssymb}
\usepackage{graphicx}
\usepackage{stackrel}
\usepackage{natbib}
\usepackage[usenames,dvipsnames]{xcolor}

\definecolor{tl}{RGB}{0,180,120}
\usepackage[colorlinks,citecolor=tl,linkcolor=red,urlcolor=magenta,unicode=true]{hyperref}

\begin{document}
	\title{Constraining PBH mass distributions from 21cm brightness temperature results and an analytical mapping between probability distribution of 21cm signal and PBH masses}
	
	\author{Upala Mukhopadhyay}
	\email{upala.mukhopadhyay@saha.ac.in}
	\affiliation{Theory Division, Saha Institute of Nuclear Physics, 1/AF, Bidhannagar, Kolkata 700064, India.\\
		Homi Bhabha National Institute, Training school complex, Anushaktinagar, Mumbai 400094, India.}
	
	\author{Debasish Majumdar}
	\email{debasish.majumdar@saha.ac.in}
	\affiliation{Theory Division, Saha Institute of Nuclear Physics, 1/AF, Bidhannagar, Kolkata 700064, India.\\
		Homi Bhabha National Institute, Training school complex, Anushaktinagar, Mumbai 400094, India.}
	
	\author{Ashadul Halder}
	\email{ashadul.halder@gmail.com}
	\affiliation{S. N. Bose National Centre for Basic Sciences,\\ JD Block, Sector III, Salt lake city, Kolkata-700106, India.}
	
	\begin{abstract}
		\begin{center}
			\large{\bf Abstract}
		\end{center}
		
		The evaporation of Primordial Black Hole (PBH) via Hawking radiation influences the evolution of Inter Galactic Medium by heating up the latter and consequently affects the 21cm signal originated from the neutral Hydrogen atoms. In this work, we have considered EDGES observational data of 21cm line corresponding to cosmic dawn era to constrain the mass and the abundance of PBHs. In this context, two different PBH mass distributions namely, power law and lognormal mass distributions are considered to estimate the effects of PBH evaporation on the 21cm brightness temperature $T_{21}$. In addition to these two mass distributions, different monochromatic masses are also considered. The impacts of Dark Matter - baryon interactions on $T_{21}$ are also considered in this work along with the influences of PBH evaporation. Furthermore, adopting different monochromatic masses for PBHs, an attempt has been made to formulate a distribution for PBH masses by associating a probability weightage of the $T_{21}$ values (at $z \sim 17.2$), within the range given by EDGES experiment, with the calculated $T_{21}$ values for each of the PBH mass values. The distribution best suited for the present purpose is found to be a combination of an error function and Owen function. Allowed contours in the parameter space of (initial PBH mass-dark matter mass) are obtained.
		
	\end{abstract}
	\maketitle
	

\section{Introduction} \label{intro} 
The 21cm hydrogen spectrum is obtained due to the transition between two hyperfine spin states ($s=0$ and 1) of neutral hydrogen atoms. Since hydrogen occupies about $75\%$ of the baryonic mass of the Universe, the 21cm hydrogen spectrum could be an important probe to Cosmos in general and cosmic processes of the dark age and reionization epoch in particular. The observational results of 21cm line is generally expressed in terms of the brightness temperature $T_{21}$ of 21cm line which is dependent on the background radio temperature (in this work the background temperature is CMB temperature $T_\gamma$), spin temperature of the hydrogen gas $T_s$ ($T_s$ is the excitation temperature of the hydrogen gas) and optical depth $\tau$ of the gas and it is defined as,
\begin{equation}
T_{21} = \frac{T_s - T_\gamma}{1+z}(1-e^{-\tau})\,\,.\label{T21}
\end{equation} 
Therefore, 21cm spectrum show absorption or emission signal if $T_s < T_\gamma$ and $T_s > T_\gamma$ respectively.

Observational outcomes of EDGES's (Experiment to Detect the Global EoR Signature) has become a remarkable probe in exploration of several unknown cosmic phenomena of the cosmic dark age. EDGES experiment reported excess absorption trough in $T_{21}$ signal corresponding to cosmic dawn epoch around $z \approx 17$. According to the standard cosmological scenario, the brightness temperature at $z \approx 17$ is obtained about $-200$ mK but EDGES experiment observed that 21cm brightness temperature during cosmic dawn is $T_{21}=-500^{+200}_{-500}$ mK at $z\approx 17$. Consequently, in order to explain the observational outcomes of EDGES, a certain amount of additional cooling is required, which essentially appears in the form of baryon-DM scattering \cite{Munoz:2015bca}. However, other heating/cooling effects such as PBH evaporation \cite{Yang:2019bkk,Yang:2020egn,Clark:2018ghm, Mittal_2022}, annihilation \cite{Basu:2020qoe,DAmico:2018sxd,Natarajan:2009bm,Liu:2018uzy} and decay of Dark Matter candidates \cite{Clark:2018ghm,Mitridate:2018iag,Halder:2021uoa} etc. may perturb the global 21-signal remarkably. In spite of its popularity, the EDGES results are suffered from controversies \cite{Hills:2018vyr, Bradley:2018eev, Tauscher:2020wso}. Recently, SARAS 3 has challenged the EDGES results with 95.3$\%$ confidence level \cite{Singh:2021mxo}. Despite of this recent controversy, in this work we use EDGES result to represent any global 21cm excess absorption signal \cite{Saha:2021pqf}.

On the other hand, Primordial Black Holes (PBHs) has been the centre of interest for decades in several aspects of astrophysics and cosmology. PBHs are believed to be generated as an outcome of the collapse of the over density regions in the early epochs of the Universe \cite{Garcia-Bellido:2017fdg,Khlopov:2008qy,Belotsky:2014kca,Belotsky:2018wph}. However, besides the standard scenarios, there are few alternative conjectures for addressing the formation of PBHs, namely, collapse of domain walls and cosmic strings \cite{Hogan:1984zb,Polnarev:1988dh,Maeda:1981gw}, fragmentation of scalar condensation \cite{Cotner:2016cvr,Cotner:2018vug,Cotner:2017tir}, confinement of quark pairs (which are first pushed apart by
inflation and then confined after re-entering the horizon) \cite{Dvali:2021byy} etc. PBHs may be substantially smaller in mass in comparison to the stellar-mass black holes \cite{Yang:2020egn,Clark:2018ghm,Cang:2021owu,Saha:2021pqf}. As a result, the Hawking radiation \cite{Hawking:1974rv} from such black holes (BHs) are significantly prominent. Therefore, the phenomenon of PBH evaporation via Hawking radiation may be a promising tool in the exploration of several aspects of this hypothetical candidate of the black hole. The emitted particles in the form of Hawking radiation heat up the baryonic medium and thus could influence the global 21cm signature. 

In the present analysis, we attempt to study the mass distribution of PBHs in the context of the global 21cm signal. As it is suggested from recent studies that extended mass distribution of PBH could be favourable, we essentially focus on two special types of distribution function namely, power-law mass distribution \cite{Carr:2017jsz, Mukhopadhyay:2021puu} and lognormal mass distribution \cite{Carr:2017jsz, Mukhopadhyay:2021puu} and find constraints on the distribution parameters using the observed limit of EDGES experiment $\left(-500^{+200}_{-500}\,{\rm mK}\right)$. 

It should be mentioned here that in this work the effects of Dark Matter - baryon (DM - baryon) scattering on the 21cm spectrum are also taken into account.
The cooling effect due to baryon - DM interaction essentially governed by the baryon - DM scattering cross-section, given by $\bar{\sigma}=\sigma_0 (v/c)^n$, where $v$ is the velocity and $c$ denotes the velocity of light in space. The index $n$ in the above expression depends on different physical processes. In the case of DM candidates having magnetic dipole moment, the index $n$ is considered as $n=+2,-2$ and $n=2,1,0,-1$ while the scattering in presence of Yukawa potential \cite{Buckley:2009in} is considered. On the other hand, for millicharged DM candidate \cite{Holdom:1985ag,Chun:2010ve}, $n=-4$ is chosen. The variation of baryon-DM cross-section is addressed in Ref.~\cite{Dvorkin:2020xga} for a wide mass range of DM. Similar studies have also been carried out in Ref.~\cite{Nadler:2019zrb,Bhoonah:2018wmw,Kovetz:2018zan}. In the present analysis, cross-section $\bar{\sigma}$ is parameterized as $\bar{\sigma}=\sigma_0 (v/c)^{-4}$ \cite{Munoz:2015bca,Mukhopadhyay:2020bml,Barkana:2018lgd} where the term $\sigma_0$ is the absolute scattering cross-section of DM - baryon scattering. 
Several recent phenomenological studies on the global 21cm signal also suggest the similar velocity dependence ($n=-4$) of the baryon-DM cross-section \cite{Munoz:2015bca,Bhoonah:2018wmw,Kovetz:2018zan,Mahdawi:2018euy,Barkana:2018lgd}. However in this work we consider $\sigma_0 \sim 10^{-41} \rm{cm^2}$. 

Along with constraining the parameters of two possible mass distributions of PBHs (power law and lognormal) with EDGES like results, we also propose a distribution of PBH mass based on the EDGES observation. In this particular case, we evaluate the weight factor at every PBH mass by computing the brightness temperature of 21cm hydrogen line and comparing that with the EDGES result. 

We discuss the imprints of PBH on 21cm line in section \ref{sect. PBH} while in section \ref{sect. DM} the impacts of DM - baryon scattering on the spectrum are discussed and temperature evolutions are briefly described in section \ref{sect. T21}.  The calculations and results are furnished in section \ref{results} and finally in section \ref{summary} summary and discussions are given.
\section{Imprints of Primordial Black Holes on the 21cm Line}\label{sect. PBH}
In this section, we give a brief account of the energy injection of PBHs due to Hawking radiation on IGM. This can affect the evolution of IGM and thus the 21cm brightness signal.

\subsection{Energy Injection Effects of Primordial Black Holes on IGM}
Evaporation of PBHs through Hawking radiation can be a possible steady source of electrons/positrons and photons. These particles, emitted from relatively low mass PBHs ($M_{\rm BH} < 10^{15}$ g), can interact with IGM and thus consequently modify the 21cm brightness temperature \cite{Mack:2008nv, Clark:2018ghm, Mittal_2022, Yang:2020egn}.

The rate of mass loss of a PBH having mass $M_{\rm BH}$ due to Hawking radiation is given by \cite{Hawking:1974rv,Clark:2018ghm},
\begin{equation}
\frac{d M_{\rm BH}}{dt}\approx -5.34 \times 10^{25} \sum_i \phi_i \left(\frac{M_{\rm BH}}{{\rm g}}\right)^{-2} {\rm g}\, {\rm sec^{-1}}\,\,, \label{mdecay}
\end{equation}
where the coefficient $\phi_i$ denotes the evaporation fraction of the $i$-th particle. The total evaporation fraction is calculated as \cite{MacGibbon:1991tj},

\begin{eqnarray}
\sum_i \phi_i = && 1.569+0.569{\rm exp}\left(-\frac{0.0234}{T_{\rm PBH}}\right)+3.414{\rm exp}\left(-\frac{0.066}{T_{\rm PBH}}\right)+1.707{\rm exp}\left(-\frac{0.11}{T_{\rm PBH}}\right) \nonumber \\
&& +0.569{\rm exp}\left(-\frac{0.394}{T_{\rm PBH}}\right)+1.707{\rm exp}\left(-\frac{0.413}{T_{\rm PBH}}\right)+1.707{\rm exp}\left(-\frac{1.17}{T_{\rm PBH}}\right)\nonumber\\
&&+1.707{\rm exp}\left(-\frac{22}{T_{\rm PBH}}\right)+0.963{\rm exp}\left(-\frac{0.1}{T_{\rm PBH}}\right)\,\,.
\end{eqnarray}
Therefore, the total evaporation rate is dependent on the temperature of PBH ($T_{\rm PBH}$) which is defined as $T_{\rm PBH} \approx 1.06 \times \left(\frac{10^{13} {\rm g}}{M_{\rm BH}}\right)$ GeV \cite{Hawking:1974rv}.

The energy injection rate per unit volume due to the evaporation of PBH is computed as \cite{MacGibbon:1991tj},

\begin{equation}
	\left.\frac{dE}{dVdt}\right|_{\rm PBH} = - \frac{dM_{\rm BH}}{dt}n_{\rm PBH}(z)\,\,, \label{PBH_energy}
\end{equation}

where $n_{\rm PBH}(z)$ is the number density of PBHs at redshift $z$ and defined by \cite{Yang:2020egn},

\begin{equation}
n_{\rm PBH}(z) \approx 1.46 \times 10^{-4} \beta_{\rm BH} (1+z)^3 \left(\frac{M_{{\rm BH},i}}{{\rm g}}\right)^{-3/2}\,\rm{cm^{-3}}\,.
\end{equation}

In the above equation, $M_{{\rm BH},i}$ represents the initial mass of PBH and $\beta_{\rm BH}$ is the initial mass fraction of PBH.
\subsection{Mass Distribution Functions of Primordial Black Holes}
In order to compute the energy injection of PBHs by using the above Eq.~\ref{PBH_energy}, it is considered that PBHs would have monochromatic mass distribution i.e., all the PBHs would be of identical masses, but some recent studies suggest that  extended mass distributions would rather be favourable \cite{Carr:2017jsz}. For such extended mass functions of PBHs the energy injection rate per unit volume is calculated as,

\begin{equation}
\left.\frac{dE}{dVdt}\right|_{\rm total} =\dfrac{\int_{M_{\rm min}}^{M_{\rm max}} dM_{{\rm BH}} \,g(M_{{\rm BH}}) \left.\frac{dE}{dVdt}\right|_{\rm PBH}}{\int_{M_{\rm min}}^{M_{\rm max}} dM_{{\rm BH}} \,g(M_{{\rm BH}})}\,\,, \label{energy_extended}
\end{equation}
with $g(M_{{\rm BH}})$ being the mass distribution function of PBHs. In the above expression $M_{\rm min}$ and $M_{\rm max}$ are the minimum and maximum value of the chosen mass spectrum of PBHs. Hence, in this work we have considered two theoretically motivated mass distribution functions of PBHs namely, power law mass distribution \cite{Carr:2017jsz} and lognormal mass distribution \cite{Carr:2017jsz}. 

In order to simplify our calculation, we break the entire mass distribution into 500 bins in logarithmic scale within the chosen mass range. The masses of the midpoints of each bins are $M_{{\rm BH},i}$, $i=1,2,3,...,500$. So now Eq.~\ref{energy_extended} takes the form,
\begin{equation}
	\left.\frac{dE}{dVdt}\right|_{\rm total} = \dfrac{\sum_{i=1}^{500} \,g(M_{{\rm BH},i}) \Delta M_i \left.\frac{dE}{dVdt}\right|_{{\rm PBH,\,for}M_{{\rm BH},i}}}{\sum_{i=1}^{500} \,g(M_{{\rm BH},i}) \Delta M_i}\,\,, \label{energy_extended_fnl}
\end{equation}
where $\Delta M_i$ is the bin width of the $i^{\rm th}$ bin. Note that Eq.~\ref{energy_extended_fnl} is in fact mass evolution equation of PBH (given in Eq.~\ref{mdecay}) but to be evaluated for 500 different PBH masses and then to be summed over.

Power law mass distribution of PBHs arises from the scale invariant density fluctuations or from the cosmic string collapse \cite{Carr:2017jsz} and this mass distribution function is expressed as \cite{Carr:2017jsz, Mukhopadhyay:2021puu, Chan:2020zry},
\begin{equation}
g(M_{\rm BH})=\frac{\gamma}{M_{\rm max}^\gamma - M_{\rm min}^\gamma}M_{\rm BH}^{\gamma-1}\,\,, \label{power law}
\end{equation}
where $\gamma$ represents the power law index and $M_{\rm max}$ and $M_{\rm min}$ denote maximum mass limit and minimum mass limit of PBHs respectively. The power law index ($\gamma$) is related to the equation of state ($\omega$) at the time of PBH formation with the relation  $\gamma=-\frac{2 \omega}{1+\omega}$ \cite{Grindlay:1975eb}. Since PBH formations are assumed to take place at post inflationary time, the values of power index $\gamma$ would be $\gamma \in \left\lbrace -1,1\right\rbrace$.

On the other hand, lognormal mass distribution of PBHs is considered when PBHs are formed from a smooth symmetric peak in the inflationary power spectrum \cite{Dolgov:1992pu}. The lognormal mass function is defined by \cite{Carr:2017jsz, Mukhopadhyay:2021puu, Chan:2020zry},
\begin{equation}
g(M_{\rm BH})=\frac{1}{\sqrt{2 \pi} \sigma M_{\rm BH}}{\rm exp}\left(-\frac{{\rm ln}^2(M_{\rm BH}/\mu)}{2 \sigma^2}\right)\,\,,
\label{log}
\end{equation}
while $\mu$ and $\sigma$ are respectively the mean and standard deviation of the lognormal distribution. Such mass function of PBHs is first observed in Ref. \cite{Kannike:2017bxn} to address a mechanism of PBH formation for a model of baryogenesis. Later some authors have discussed this type of PBH mass distribution both theoretically and numerically \cite{Green:2016xgy}.
\section{Impacts of Dark Matter - Baryon Interaction on the 21cm Line}\label{sect. DM}
In this work we have also considered the impacts of the interaction between Dark Matter (DM) and baryonic matter on the 21cm signal. In literature it is discussed that due to DM - baryon interaction, baryon can transfer heat to the colder DM fluid and hence influence the evolution of 21cm brightness temperature \cite{Tashiro:2014tsa, Barkana:2018lgd}. Moreover, the relative velocity ($V_{\chi b}$) between DM and baryon fluid would also affect the 21cm line \cite{Munoz:2015bca} as the tendency to damp their relative velocity will heat up both of the fluids. Hence, in this work we have considered both of the above mentioned effects of DM - baryon interaction on the 21cm line. The heating rate of baryon can be evaluated from Ref. \cite{Munoz:2015bca} as,
\begin{eqnarray}
\dfrac{d Q_b}{dt} &=& \dfrac{2m_b \rho_\chi \sigma_0 e^{-r^2/2} (T_{\chi} - T_b)}{(m_\chi + m_b)^2 \sqrt{2 \pi} u_{\rm th}^3} + \dfrac{\rho_\chi}{\rho_m} \dfrac{m_\chi m_b}{m_\chi + m_b} V_{\chi b} D(V_{\chi b})\,\,, \label{heat_baryon}
\end{eqnarray}
Here, $\rho_\chi$, $\rho_m$ are energy densities of DM and total matter respectively while $T_\chi$ and $T_b$ denote DM temperature and baryon temperature respectively. The masses of DM and baryon are represented by $m_\chi$ and $m_b$.  In the above equation, the first term on the r.h.s arises from the temperature difference of DM and baryon ($T_\chi-T_b$) and the second term originates due to the velocity difference ($V_{\chi b}$) between them. The drag term $D(V_{\chi b})$ is calculated as \cite{Munoz:2015bca},
\begin{equation}
D(V_{\chi b}) \equiv -\frac{d V_{\chi b}}{d t} = \frac{\rho_m \sigma_0}{m_b+m_\chi}\frac{1}{V^2_{\chi b}}  F(r)\,\,, \label{DV_chib}
\end{equation}
where $r \equiv V_{\chi b}/u_{\rm th}$, $u_{\rm th}^2 \equiv \frac{T_b}{m_b}+\frac{T_\chi}{m_\chi}$ and $F(r) \equiv {\rm erf}\left(\frac{r}{\sqrt{2}}\right)-\sqrt{\frac{2}{\pi}}e^{-r^2/2}r$. The parametrization $\bar{\sigma} = \sigma_0 v^{-4}$ for interaction cross section of DM and baryon fluid is considered for this calculation. The heating rate of DM ($\frac{d Q_\chi}{dt}$) can be obtained by interchanging $\chi \leftrightarrow b$ in Eq.~\ref{heat_baryon}.
\section{Temperature Evolutions and 21cm Signal}\label{sect. T21}
In this section, we calculate the evolutions of temperatures ($T_b$, $T_\chi$) and 21cm signal by including the effects of the energy injection of PBHs and DM - baryon interaction. The temperature evolutions of DM and baryon can be calculated by solving the following coupled differential equations \cite{BH_F},
\begin{eqnarray}
\frac{d T_\chi}{d z} &=& \frac{2 T_\chi}{1+z} - \frac{2 \dot{Q}_\chi}{3 (1+z) H(z)}\,\,, \label{T_chi}\\
\frac{d T_b}{d z} &=& \frac{2 T_b}{1+z} + \frac{\Gamma_c}{(1+z) H(z)}(T_b - T_\gamma)-\frac{2 \dot{Q}_b}{ 3 (1+z) H(z)} \nonumber \\
&&-\frac{2}{3k_b H(z) (1+z)}\frac{K_{\rm PBH}}{1+f_{\rm He}+x_e}\,\,. \label{T_b}
\end{eqnarray}
Here, $T_\gamma=2.725(1+z)$ K is the photon temperature and $\Gamma_c=\frac{8\sigma_T a_r T_\gamma^4 x_e}{3 (1 + f_{\rm He} +x_e) m_e c}$ denotes the Compton interaction rate where $\sigma_T$ and $a_r$ are the Thomson scattering cross section and the radiation constant respectively. The fractional abundance of He is denoted by $f_{\rm He}$ while the free electron abundance is $x_e=n_e/n_H$ ($m_e$ and $c$ is the electron mass and the speed of light). The third term on the r. h. s of Eq.~\ref{T_b} includes the effect of energy injection from PBHs due to Hawking radiation where $K_{\rm PBH}$ is expressed as \cite{Clark:2018ghm,Yang:2015cva, Chen:2003gz, Zhang:2007zzh, Mack:2008nv},
\begin{equation}
K_{\rm PBH}=\chi_h f(z) \frac{1}{n_b}\left.\frac{dE}{dVdt}\right|_{\rm total}\,\,,
\end{equation}
with $\chi_h = (1+2 x_e)/3$ being the fraction of the emitted energy contributes to the heating of IGM and the parameter $f(z)$, stands for the ratio of the total amount of deposited energy to the energy injected to the medium due to PBH evaporation \cite{corr_equs,fcz001,fcz002,fcz003,fcz004}.

To compute the evolution of baryon temperature, the evolution of free electron fraction $x_e$ is needed to be calculated simultaneously. The evolution equation of $x_e$ is expressed as \cite{Ali-Haimoud:2010hou}
\begin{equation}
\frac{d x_e}{d z} = \frac{C_P}{(1+z) H(z)}\left(n_H A_B x_e^2 - 4 (1-x_e)B_B e^{\frac{-3 E_0}{4 T_\gamma}}\right) - \frac{1}{(1+z)H(z)}I_{\rm PBH}(z)\,\,, \label{xe}
\end{equation}
where the Peebles $C$-factor \cite{Peebles:1968ja} is represented by $C_p$, $E_0$ denotes the ground state energy of Hydrogen ($E_0=13.6$ eV) while the effective recombination coefficient and the effective photoionization rate to and from the excited states are $A_B$ and $B_B$ respectively \cite{Ali-Haimoud:2010tlj}. In the above Eq.~\ref{xe}, the $I_{\rm PBH}$ denotes the ionization rate caused by the energy injection of PBHs and it is defined by \cite{Clark:2018ghm,Yang:2015cva, Chen:2003gz, Zhang:2007zzh,Mack:2008nv},

\begin{equation}
I_{\rm PBH}=\chi_i f(z) \frac{1}{n_b}\frac{1}{E_0}\left.\frac{dE}{dVdt}\right|_{\rm total}\,\,,
\end{equation}

with $\chi_i=(1-x_e)/3$ is the fraction of injected energy influences ionization of the IGM.

To obtain the evolutions of $T_b$ and $T_\chi$ the variations of the relative velocity between DM and baryon should also be simultaneously calculated with the differential equation \cite{Munoz:2015bca},
\begin{equation}
\frac{d V_{\chi b}}{dz} = \frac{V_{\chi b}}{1+z}+\frac{D(V_{\chi b})}{(1+z) H(z)}\,\,. \label{V_chib}
\end{equation}

Since Eqs.~\ref{T_chi}, \ref{T_b}, \ref{xe}, \ref{V_chib} and \ref{energy_extended_fnl} are all coupled, we need to solve these five equations simultaneously with proper initial conditions to compute the evolutions of baryon temperature $T_b$ and thus to obtain the 21cm brightness temperature. Now, the spin temperature $T_s$, defined by the ratio of the number densities of Hydrogen atoms in spin triplet and spin singlet states ($n_1/n_0 = g_1/g_0 \exp(-h\nu/kT_s)$), is calculated from the expression \cite{Pritchard:2011xb},
\begin{equation}
T_s^{-1}=\frac{T_\gamma^{-1} +y_c T_b^{-1} +y_\alpha T_\alpha^{-1}}{1 +y_c +y_\alpha}\,\,. \label{spin}
\end{equation}

Here, $y_\alpha$ and $y_c$ denote the Lyman-$\alpha$ coupling parameter and collisional coupling parameter respectively \cite{BH_21cm_2,Yuan_2010,Kuhlen_2006} while $T_\alpha$ is temperature of the Lyman-$\alpha$ background which is identical to baryon temperature $T_b$ for $z\lessapprox20$ \cite{Yang:2021idt}.

As mentioned in Sect.~\ref{intro}, the 21cm brightness temperature $T_{21}$ can now be calculated from the definition,
\begin{eqnarray}
T_{21} &=& \frac {T_s - T_\gamma} {1+z} (1 - e^{-\tau}) \simeq 
\frac{T_s - T_\gamma} {1 + z} \tau\,\,, \label{T21_1}
\end{eqnarray}
where $\tau$ is the optical depth expressed as $\tau=\frac{3}{32\pi}\frac{T_*}{T_{\rm s}}n_{\rm HI}\lambda_{21}^3 \frac{A_{10}}{H(z)}$ \cite{Pritchard:2011xb} (here $A_{10}$ signifies the Einstein coefficient for spontaneous emission due to the transition from triplet to singlet state \cite{Ali-Haimoud:2010tlj,Ali-Haimoud:2010hou}, wavelength of 21cm line is denoted by $\lambda_{21}$ while $n_{\rm HI}$ is the number density of neutral Hydrogen and
$ T_*$ represents the 21cm photon transition temperature).

\section{Calculations and Results}\label{results}
In this section, we describe our calculations and results using the formalism described in Sect.~\ref{sect. PBH} -~\ref{sect. T21}. Evolutions of $T_s$, $T_b$ and $T_{21}$ are calculated for two mass distributions of PBH (lognormal mass distribution and power law mass distribution) and bounds on model parameters are computed in this context with EDGES like observational limit. Constraints on DM mass $m_\chi$, initial mass fraction of PBH $\beta_{\rm BH}$ and parameters of PBH mass distributions are estimated for the above mentioned cases with EDGES's results. Moreover, a  mass distribution function of PBH is derived in such a way that it can predict the EDGES limit $T_{21}=-500^{+200}_{-500}$ mK at reionization epoch (discussed is Subsect.~\ref{ssec:edges_dist_asha}).
\subsection{Lognormal Mass Distribution of Primordial Black Hole}\label{ssec:lognormal}
\begin{figure}
\centering
\begin{tabular}{cc}
\includegraphics[width=0.5\linewidth]{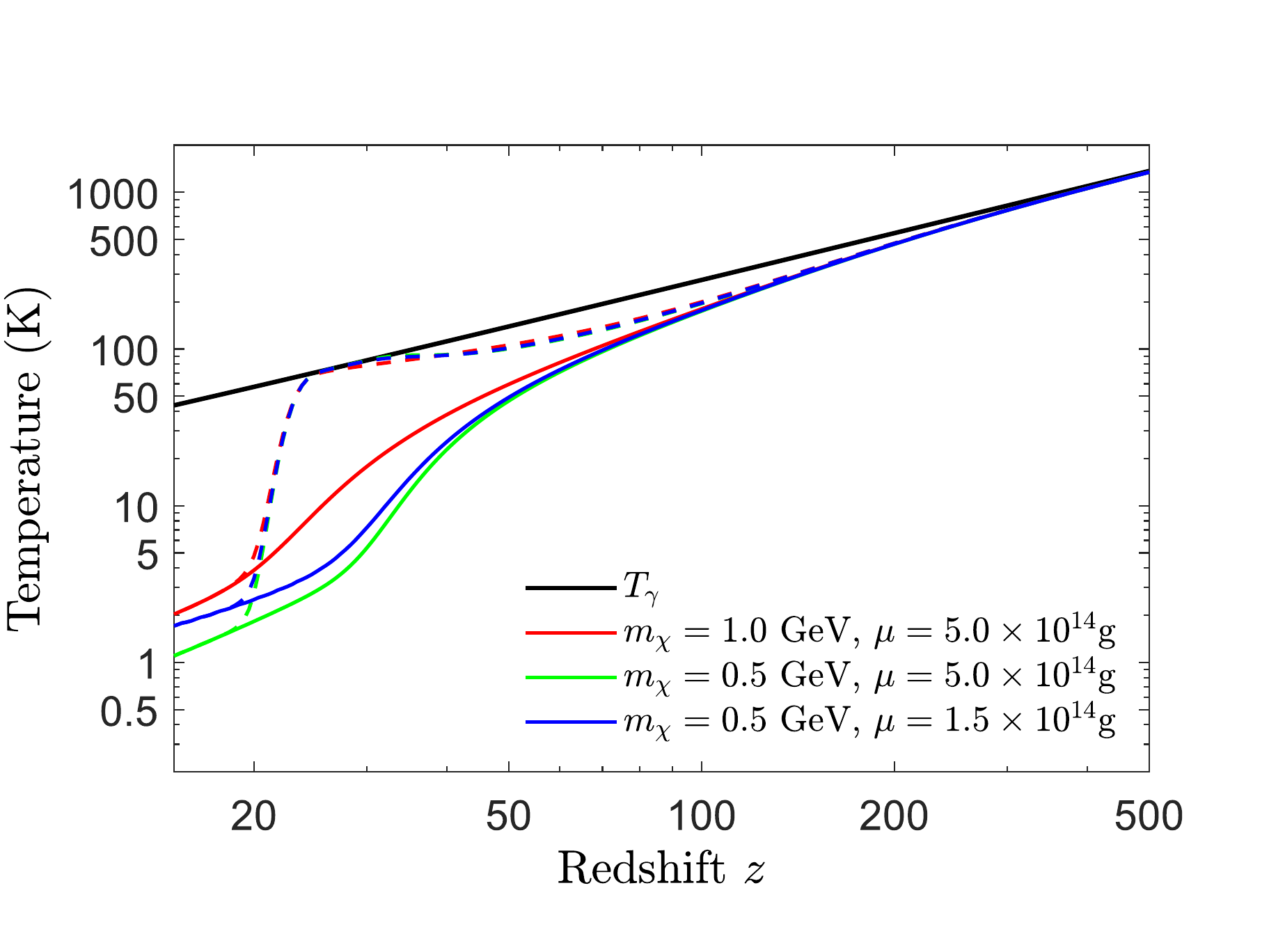}&
\includegraphics[width=0.5\linewidth]{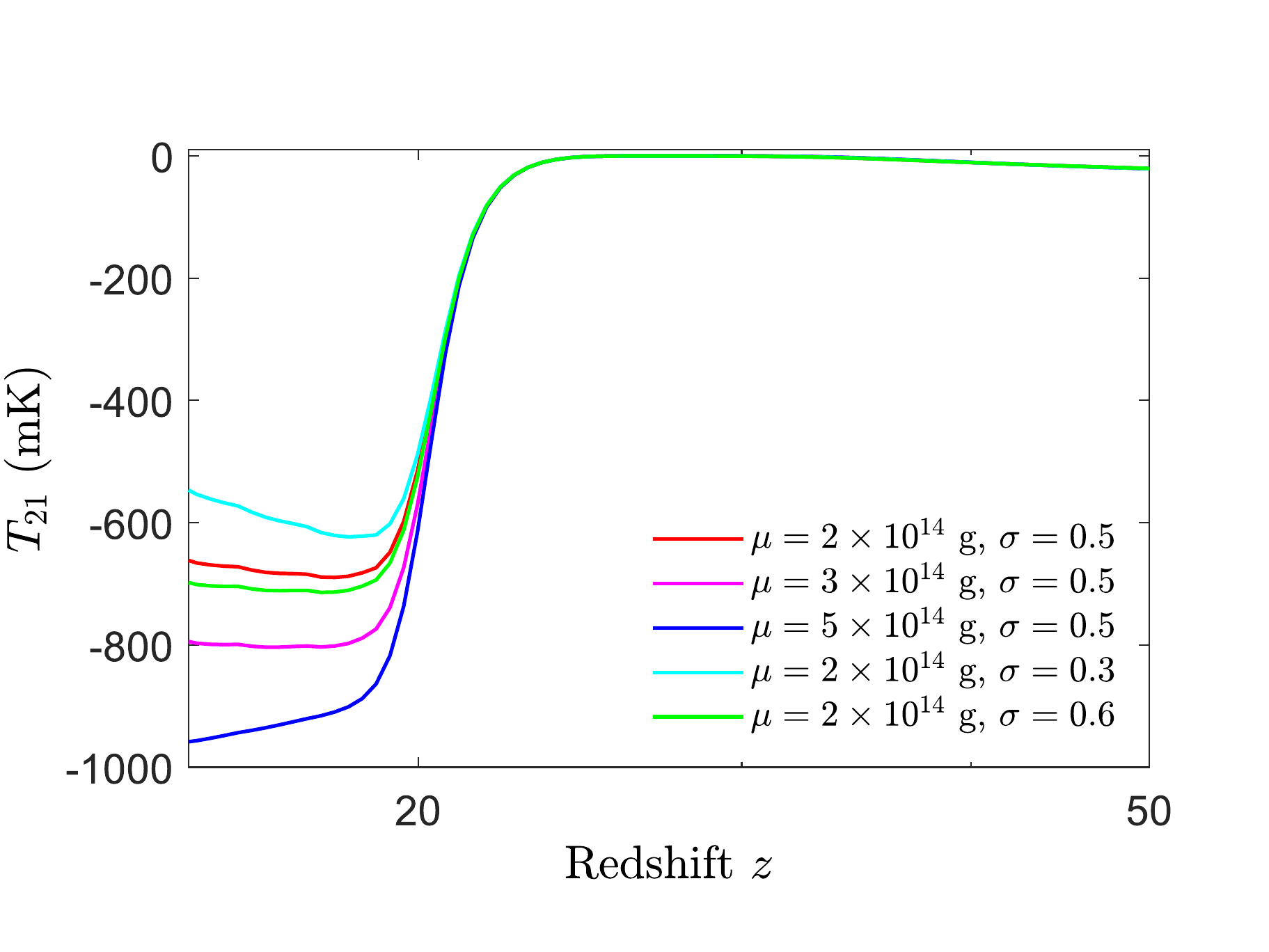}\\
(a)&(b)\\
\end{tabular}
\caption{ \label{fig:temp} (a) Evolutions of $T_s$ (dashed lines) and $T_b$ (solid lines) with $z$ for different mean values $\mu$ of lognormal distribution of PBH and different DM mass $m_\chi$. (b) variations of $T_{21}$ with $z$ for different mean $\mu$ and variance $\sigma$ values of the mass distribution of PBH.}
\end{figure}

Lognormal distribution of PBH masses is considered in this section to study the effects of PBH energy injections and DM - baryon interactions on 21cm brightness temperature. In Fig.~\ref{fig:temp}(a) evolutions of $T_b$ (solid lines in the plot) and corresponding spin temperature $T_s$ (dashed lines in the plot) with redshift $z$ are plotted for different values of DM mass ($m_\chi$= 0.5 GeV, 1 GeV) and different mean values of the distribution ($\mu$= 5$\times 10^{14}$ g and 1.5$\times 10^{14}$ g). It can be noted from Fig.~\ref{fig:temp}(a) that a smaller gas temperature $T_b$ (and $T_s$) is obtained at the reionization epoch when a larger value of $\mu$ ($\mu$= 5.0$\times 10^{14}$ g) is considered for a fixed value of $m_\chi$ ($m_\chi=$ 0.5 GeV). It indicates the fact that the energy injection rates of PBHs with smaller masses are higher than the same with larger masses. Hence, PBH mass distribution with a lower mean value can inject larger amount of energy in the IGM compared to the distribution with a higher $\mu$. From Fig.~\ref{fig:temp}(a) it can also be observed that IGM temperature decreases with the decrease of $m_\chi$. This is however expected from Eq.~\ref{heat_baryon} since the cooling rate of the baryon is inversely proportional to DM mass \cite{Mukhopadhyay:2020bml}.

Similar comments can be made from Fig.~\ref{fig:temp}(b) where the variations of 21cm brightness temperature $T_{21}$ with $z$ are shown for different mean $\mu$ and variance $\sigma$ values. Here also it can be observed that more negative values of $T_{21}$ can be obtained for larger values of $\mu$ as energy injection rates are smaller for heavier PBHs. It can also be observed from Fig.~\ref{fig:temp}(b) that larger $T_{21}$ is obtained when $\sigma=0.3$ is considered (indigo line in the Fig.~\ref{fig:temp}(b)) when compared with the same with $\sigma=0.6$ (green line in Fig.~\ref{fig:temp}(b)) where the mean value is fixed at $\mu = 2 \times 10^{14}$ g. It can be mentioned here that we have repeated the calculation by fixing $\mu = 10^{15}$ g and have found that smaller $T_{21}$ is obtained for $\sigma=0.3$ than when $\sigma=0.6$ is considered. This indicates that PBHs with smaller masses  $\lesssim 10^{14}$ g evaporate before $z \sim 17.2$ and thus contributions for lighter PBHs are not significant. But for $\mu = 10^{15}$ g,
the distribution with larger variance $\sigma=0.6$ includes larger range of PBH masses with significant contributions from the smaller masses. It may be mentioned here that black hole evaporation calculations include correction terms before full evaporation.
 
\begin{figure}
\centering
	\begin{tabular}{cc}
		\includegraphics[width=0.5\textwidth]{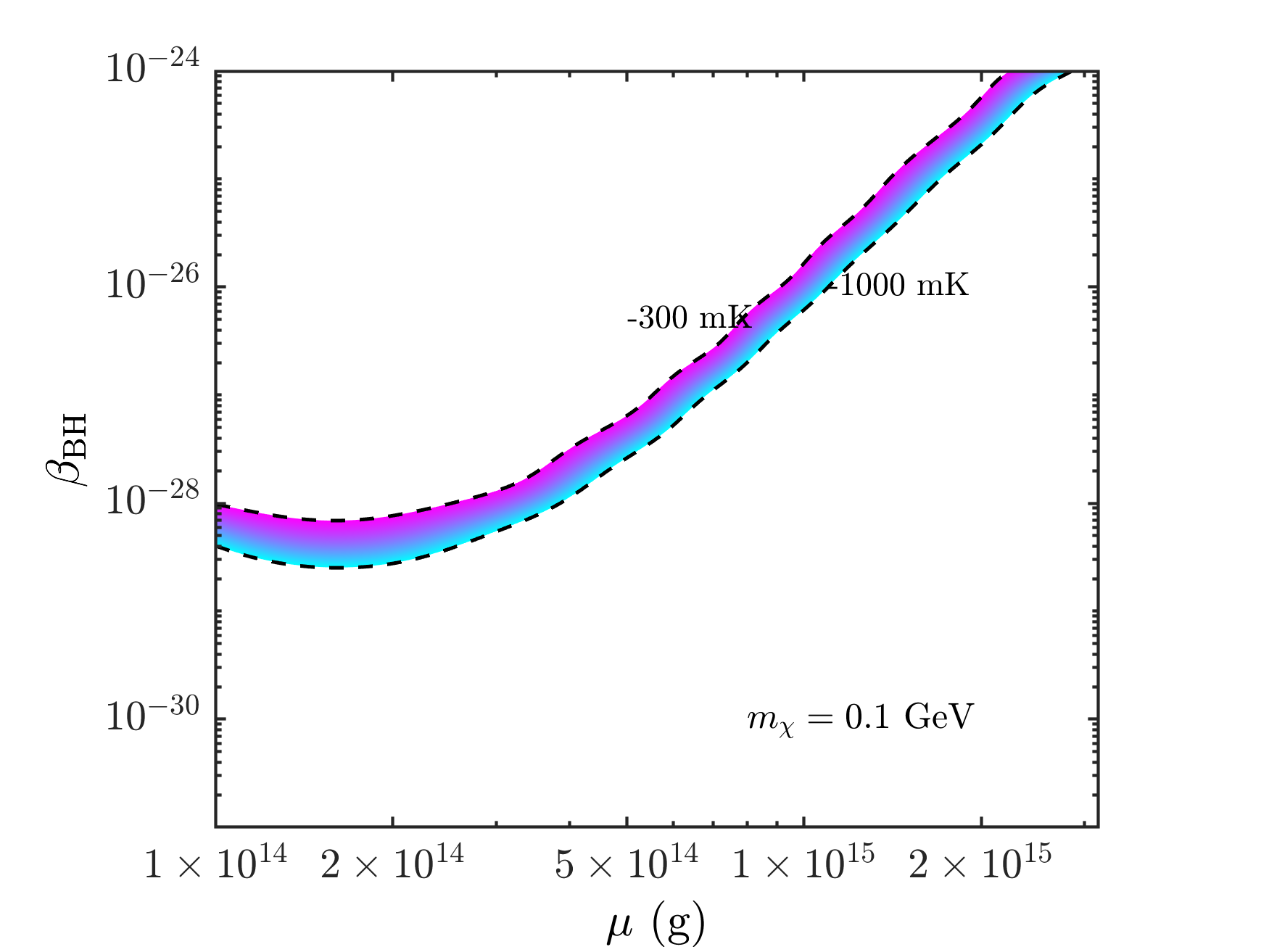}&
		\includegraphics[width=0.5\textwidth]{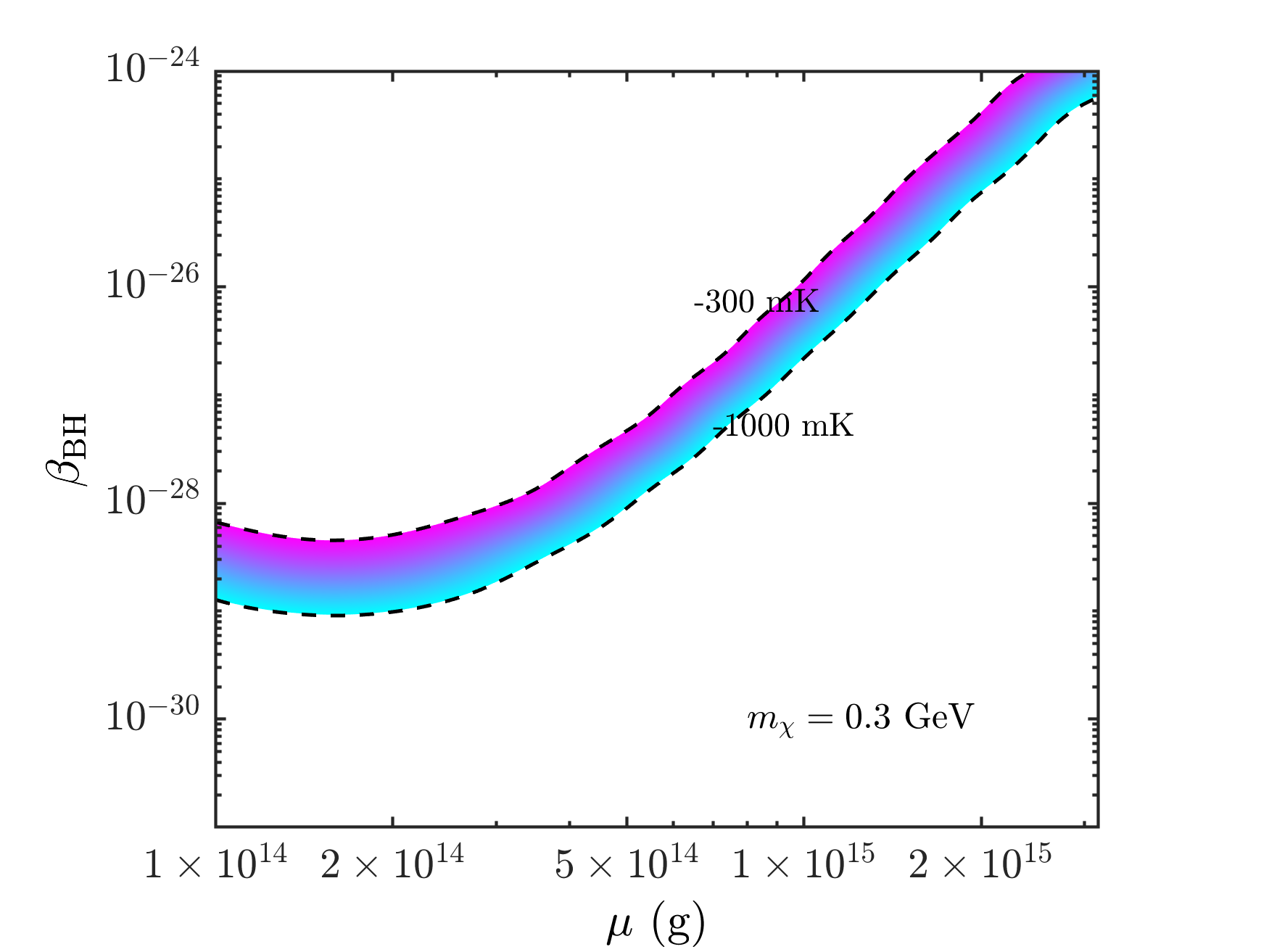}\\
		(a)&(b)\\
		\includegraphics[width=0.5\textwidth]{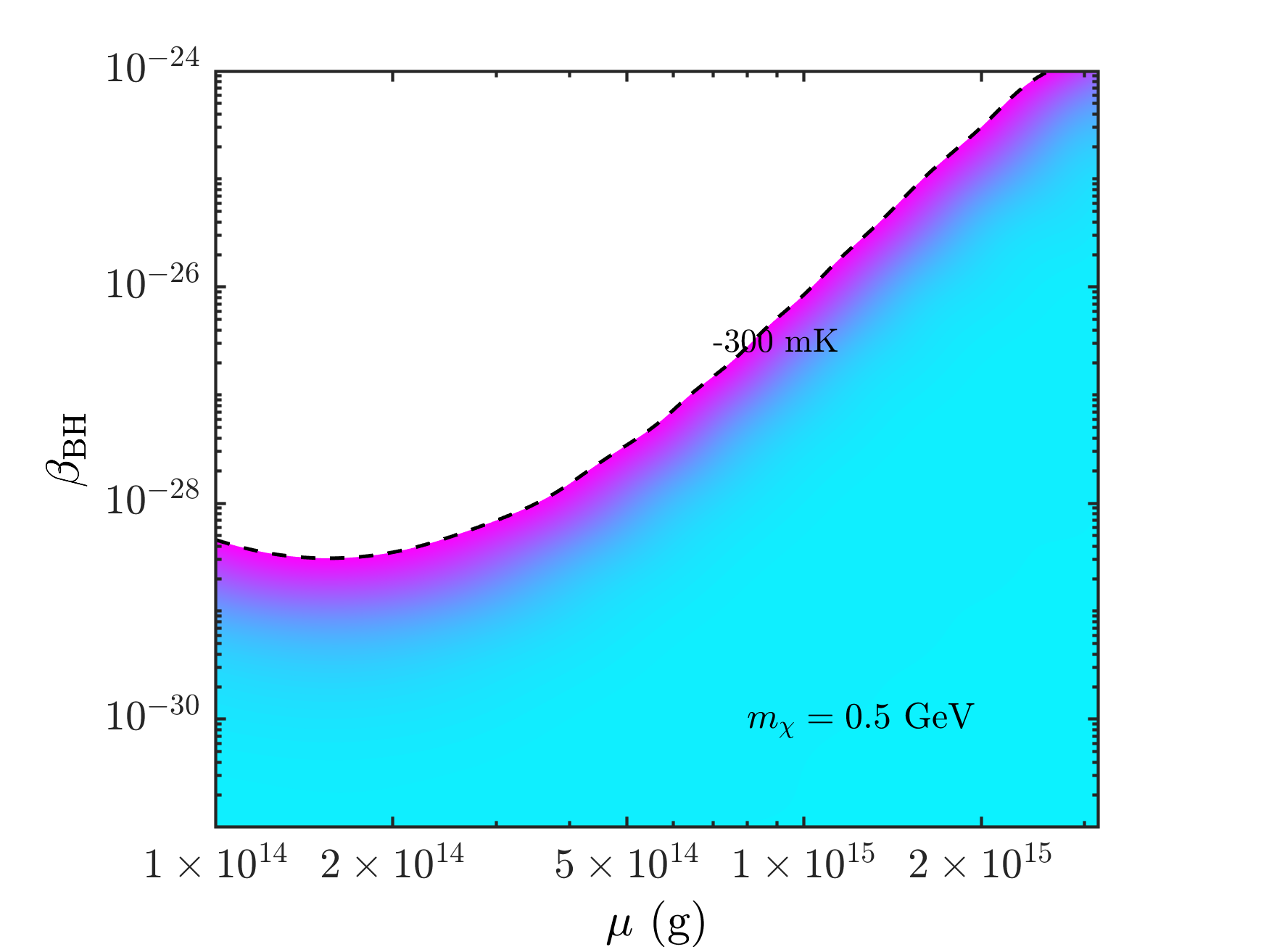}&
		\includegraphics[width=0.5\textwidth]{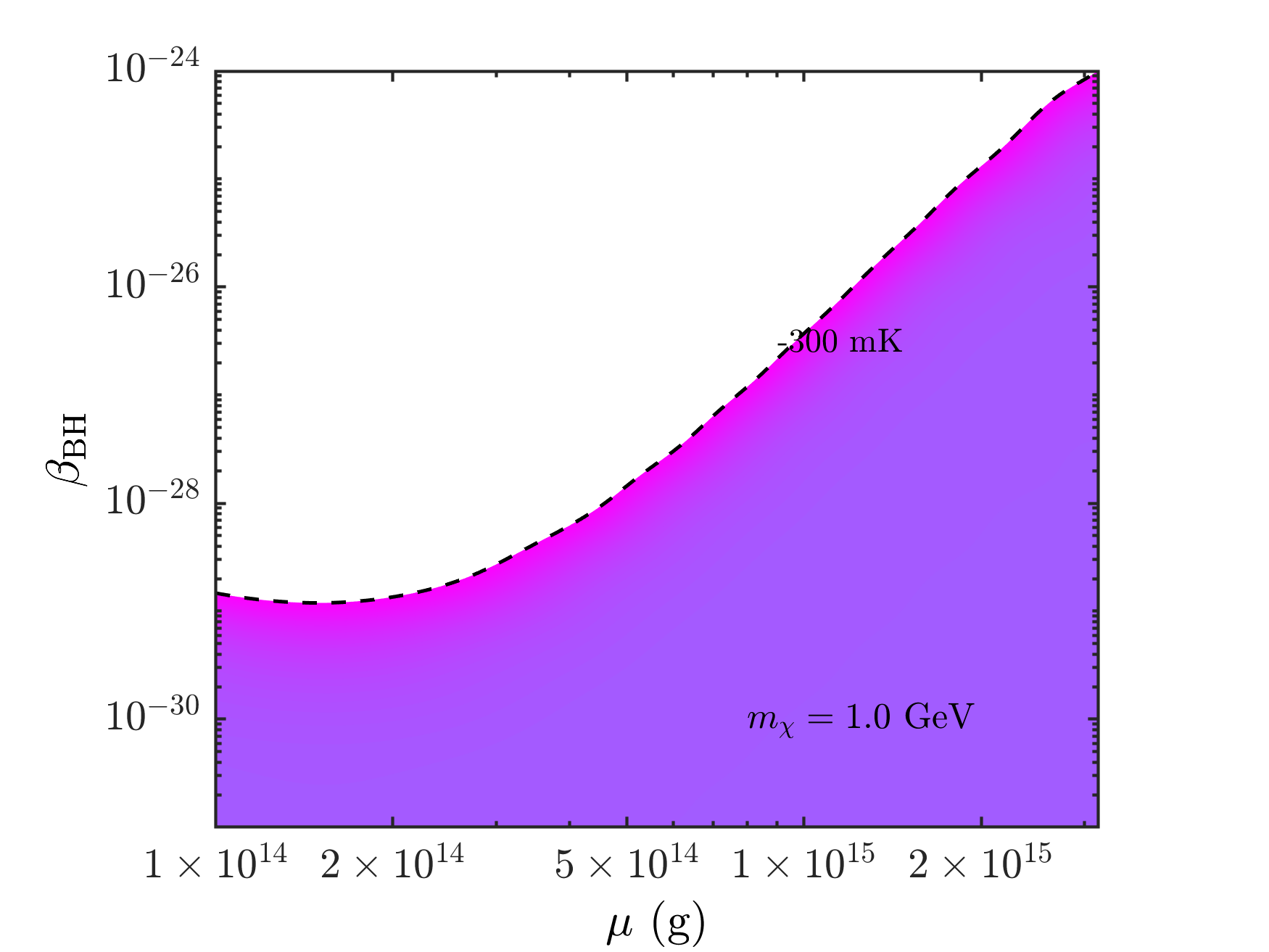}\\
		(c)&(d)\\
	\end{tabular}
	\begin{tabular}{c}
		\includegraphics[trim=0 10 0 310,clip, width=0.7\textwidth]{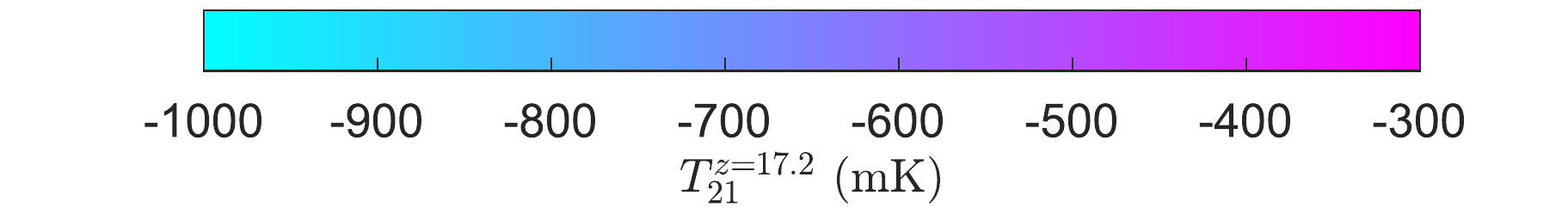}\\
	\end{tabular}
	\caption{\label{fig:beta_mu} Allowed zones in the $\beta_{\rm BH} - \mu$ plane for lognormal distribution, where different values of DM masses are considered ((a) $m_{\chi}=0.1$ GeV, (b) $m_{\chi}=0.3$ GeV, (c) $m_{\chi}=0.5$ GeV and (d) $m_{\chi}=1.0$ GeV. In all the plots, $\sigma=0.5$ has been adopted. See text for detail.}
\end{figure}

\begin{figure}
	\centering
	\begin{tabular}{cc}
		\includegraphics[trim=0 22 0 0, clip, width=0.5\textwidth]{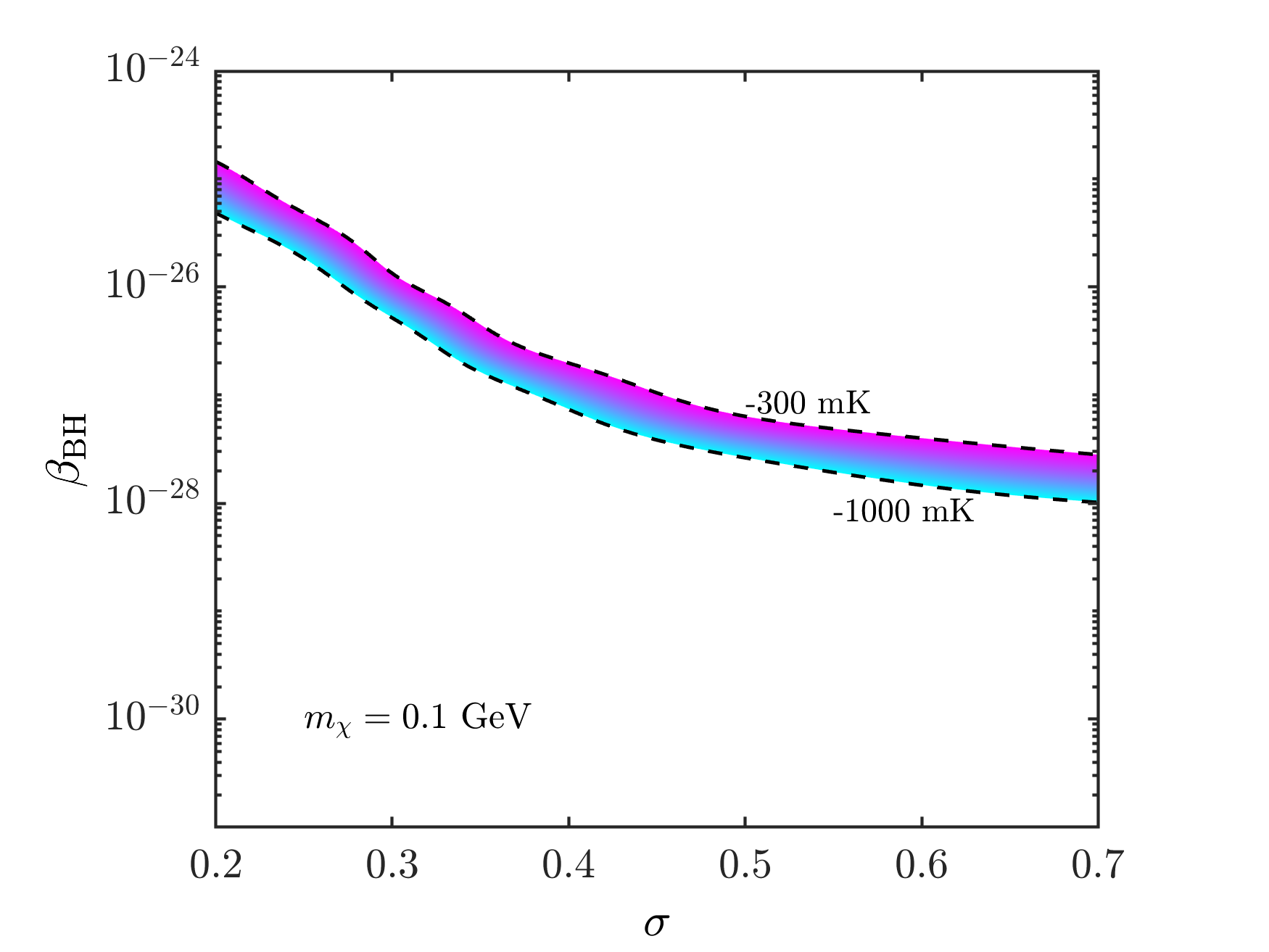}&
		\includegraphics[trim=0 22 0 0, clip, width=0.5\textwidth]{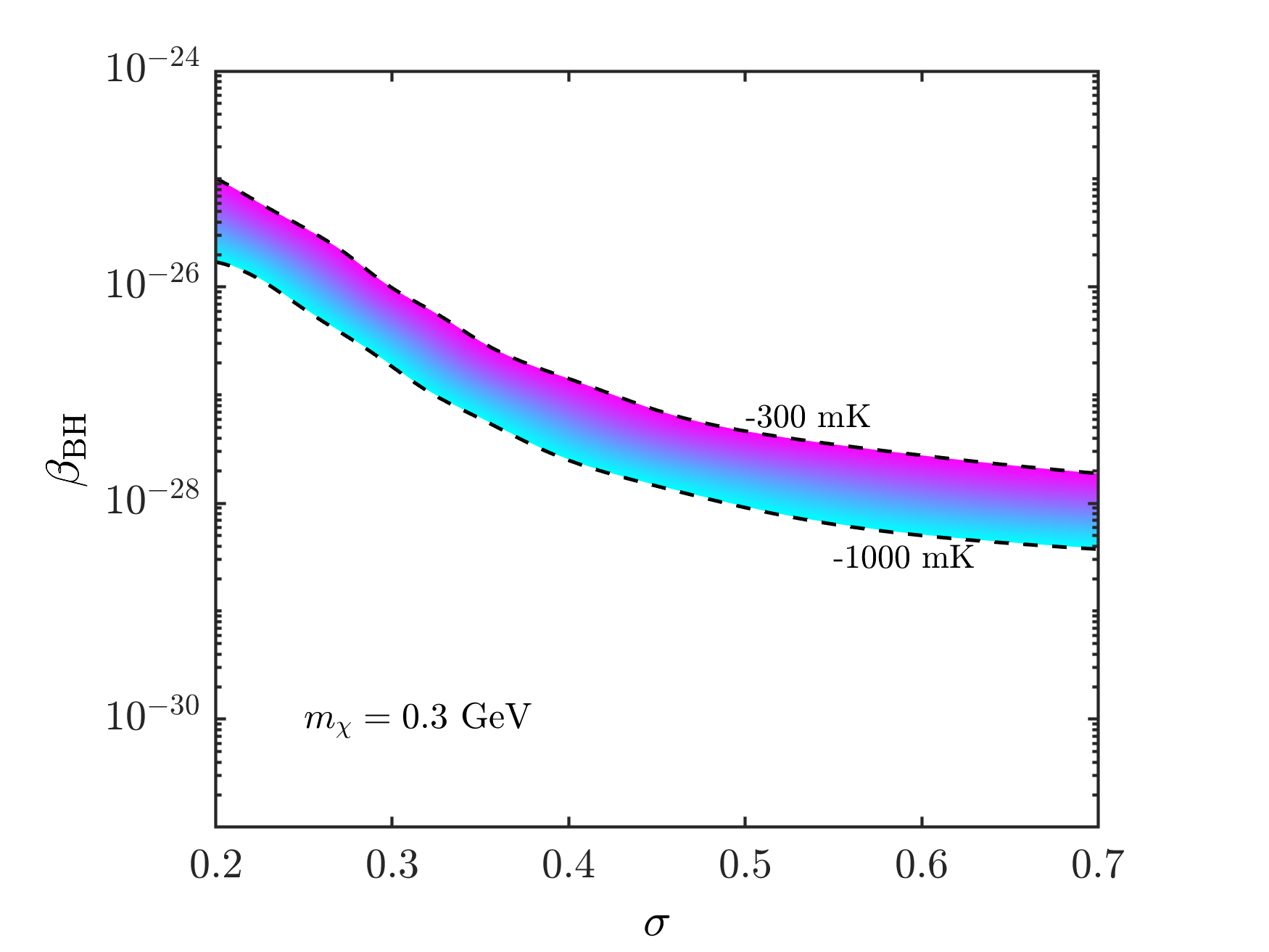}\\
		(a)&(b)\\
		\includegraphics[trim=0 22 0 0, clip, width=0.5\textwidth]{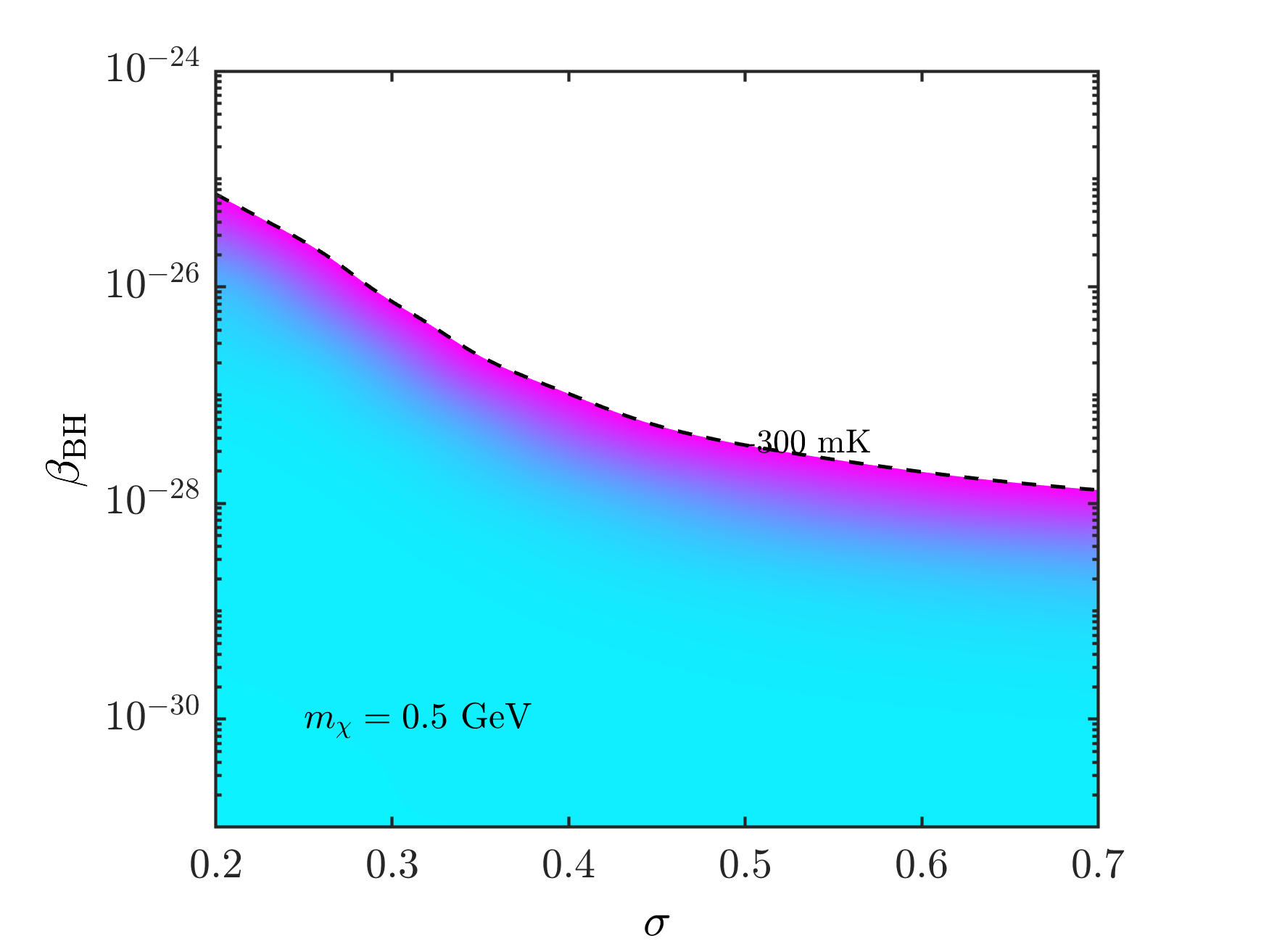}&
		\includegraphics[trim=0 22 0 0, clip, width=0.5\textwidth]{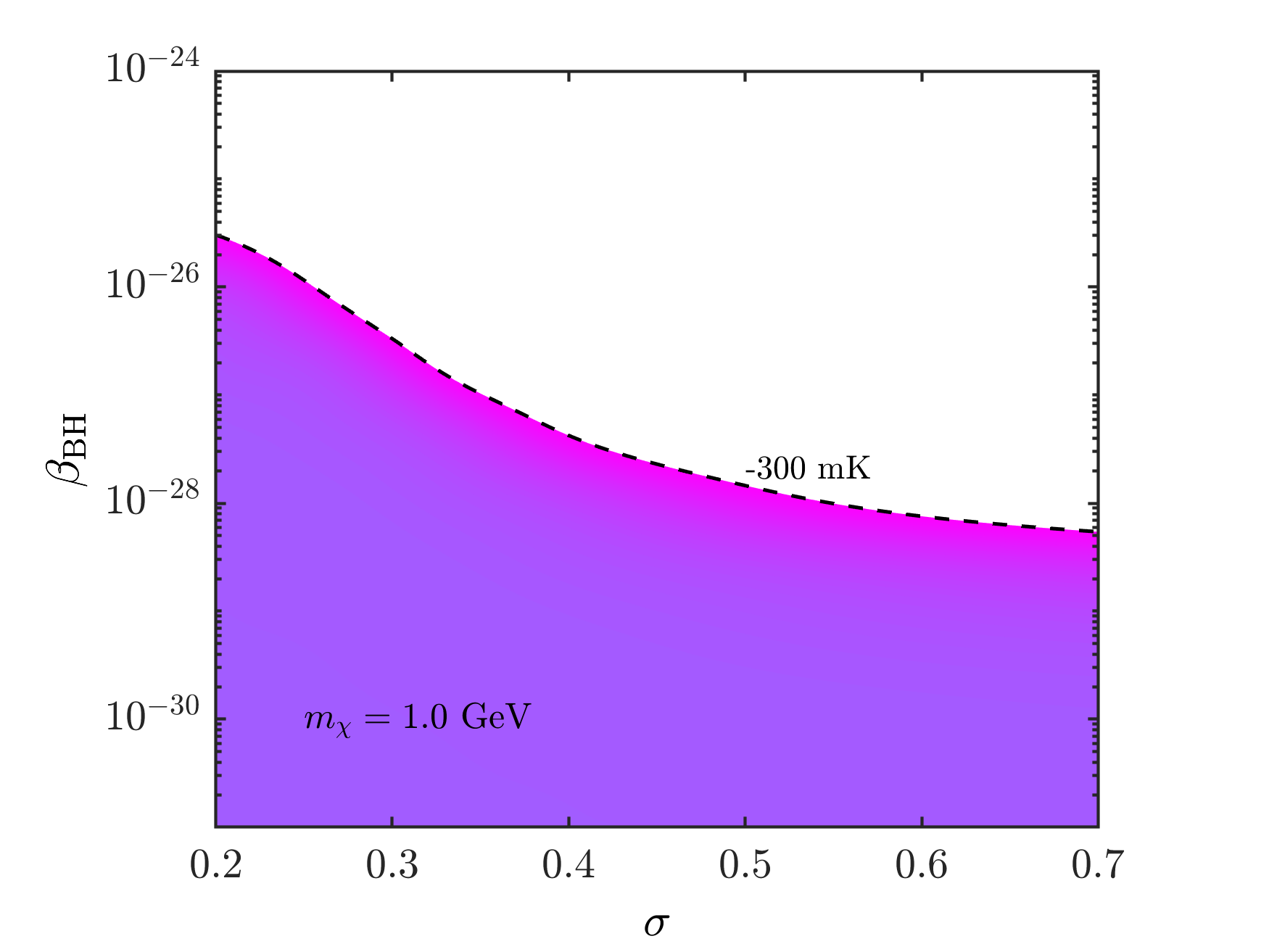}\\
		(c)&(d)\\
	\end{tabular}
	\begin{tabular}{c}
		\includegraphics[trim=0 10 0 310,clip, width=0.7\textwidth]{figure/cbar.pdf}\\
	\end{tabular}
	\caption{\label{fig:beta_sigma}Allowed zones in the $\beta_{\rm BH} - \sigma$ plane for lognormal distribution of PBH, where different values of DM masses are considered ((a) $m_{\chi}=0.1$ GeV, (b) $m_{\chi}=0.3$ GeV, (c) $m_{\chi}=0.5$ GeV and (d) $m_{\chi}=1.0$ GeV). In all the plots, $\mu=5\times 10^{14}$ g has been adopted.}
\end{figure}

One of our main focuses of the current work is to provide bounds on the parameters of PBH mass distributions and on the initial mass fraction of PBH $\beta_{\rm BH}$. In Fig.~\ref{fig:beta_mu} and Fig.~\ref{fig:beta_sigma} the allowed regions of $\beta_{\rm BH} - \mu$ plane and $\beta_{\rm BH} - \sigma$ plane are shown for different values of $m_\chi$ ((a) $m_{\chi}=0.1$ GeV, (b) $m_{\chi}=0.3$ GeV, (c) $m_{\chi}=0.5$ GeV and (d) $m_{\chi}=1.0$ GeV) by considering EDGES observational results. Upper bounds and lower bounds of $\beta_{\rm BH}$, $\mu$ and $\sigma$ are estimated by using the EDGES limit on brightness temperature of 21cm line of reionization epoch i.e., at $z \simeq 17.2$ value of $T_{21}$ is $T_{21}=-500^{+200}_{-500}$ mK. We compare our calculated values of $T_{21}$ at $z \simeq 17.2$ with EDGES limit $\left(T_{21}=-500^{+200}_{-500}\,{\rm mK} \right)$ to compute the constraints on the parameters and hence show the calculated $T_{21}$ at $z \simeq 17.2$ with a different notation $T_{21}^{z=17.2}$ with colour bars in Figs.~\ref{fig:beta_mu},~\ref{fig:beta_sigma}. It is clear from Fig.~\ref{fig:beta_mu} that larger initial mass fractions of PBH can be probed for mass distributions with higher mean values. It is expected as energy injection rate of PBH is inversely proportional to their mass value and hence larger abundance of heavier PBHs are still compatible with the EDGES results. It can also be noted from the figure that $\beta_{\rm BH}$ decreases with the decrement of $\mu$ up to a certain value $\mu \sim 2 \times 10^{14}$ g and then it starts to slightly increase. This is showing that PBHs with
mass less than $\sim 2 \times 10^{14}$ g evaporate before $z\lessapprox20$ and thus their contributions in IGM heating at $z\lessapprox20$ are comparatively lower. From Fig.~\ref{fig:beta_mu} it is also noted that for lower values of $m_\chi$ the allowed region (the coloured region showing the allowed range between the  upper limits and lower limits of the parameters) in $\beta_{\rm BH} - \mu$ is very narrow but the lower limit increases significantly with the increment of DM mass while the upper limit varies very slightly with $m_{\chi}$. Similar constraints on $\beta_{\rm BH} - M_{\rm BH}$ plane are observed for monochromatic mass distribution of PBH in Ref.~\cite{Halder:2021rbq}. For smaller $m_\chi$ ($m_\chi$ = 0.1 GeV, 0.3 GeV) the effects of DM - baryon interaction on $T_{21}$ are very high and hence $T_{21}$ fall beyond the EDGES's lower limit (less than -1000 mK) and consequently the lower limits of $\beta_{\rm BH} - \mu$ plane become stringent. As larger DM - baryon interaction rate can be acquired for smaller $m_\chi$, larger PBH abundance can be probed in these cases and hence the upper limit slightly decreases when $m_\chi$ increases.
Similar conclusion can be drawn from Fig.~\ref{fig:beta_sigma} that allowed regions in $\beta_{\rm BH} - \sigma$ plane increase with mass of DM $m_\chi$. In Fig.~\ref{fig:beta_sigma} it can be noted that larger values of $\beta_{\rm BH}$ are allowed at smaller values of $\sigma$. This is expected because larger variance ($\sigma$) indicates the inclusion of larger mass range of PBHs with lower mass values and higher evaporation rate. It can be mentioned that DM - baryon interaction cross section $\sigma_{41}$ (in the unit of $10^{-41}$ cm$^2$) is kept at $\sigma_{41}=1$ for the plots in Fig.~\ref{fig:beta_mu} and Fig.~\ref{fig:beta_sigma}. While in Fig.~\ref{fig:beta_mu} the value of $\sigma$ is fixed at $\sigma=0.5$, in  Fig.~\ref{fig:beta_sigma} the value of $\mu$ is fixed at  $\mu=5 \times 10^{14}$ g. The same computations are repeated for other values of $\sigma_{41}$ (say for $\sigma_{41}=5$) and extended allowed ranges of the parameters are obtained for larger $\sigma_{41}$ values. 

\subsection{Power Law Mass Distribution of Primordial Black Hole}\label{ssec:powerlaw}
In this section, upper and lower limits of initial mass fraction of PBH $\beta_{\rm BH}$, power index $\gamma$ and DM mass $m_\chi$ are obtained from EDGES experimental results by considering the power law mass distribution of PBH.
This is to be mentioned here that we have adopted $M_{\rm min}=6\times 10^{13}$ g and $M_{\rm max}=10^{15}$ g in Eq.~\ref{power law} (Power law distribution) for all the calculations performed in this work.

\begin{figure}
	\centering
	\begin{tabular}{cc}
		\includegraphics[width=0.5\textwidth]{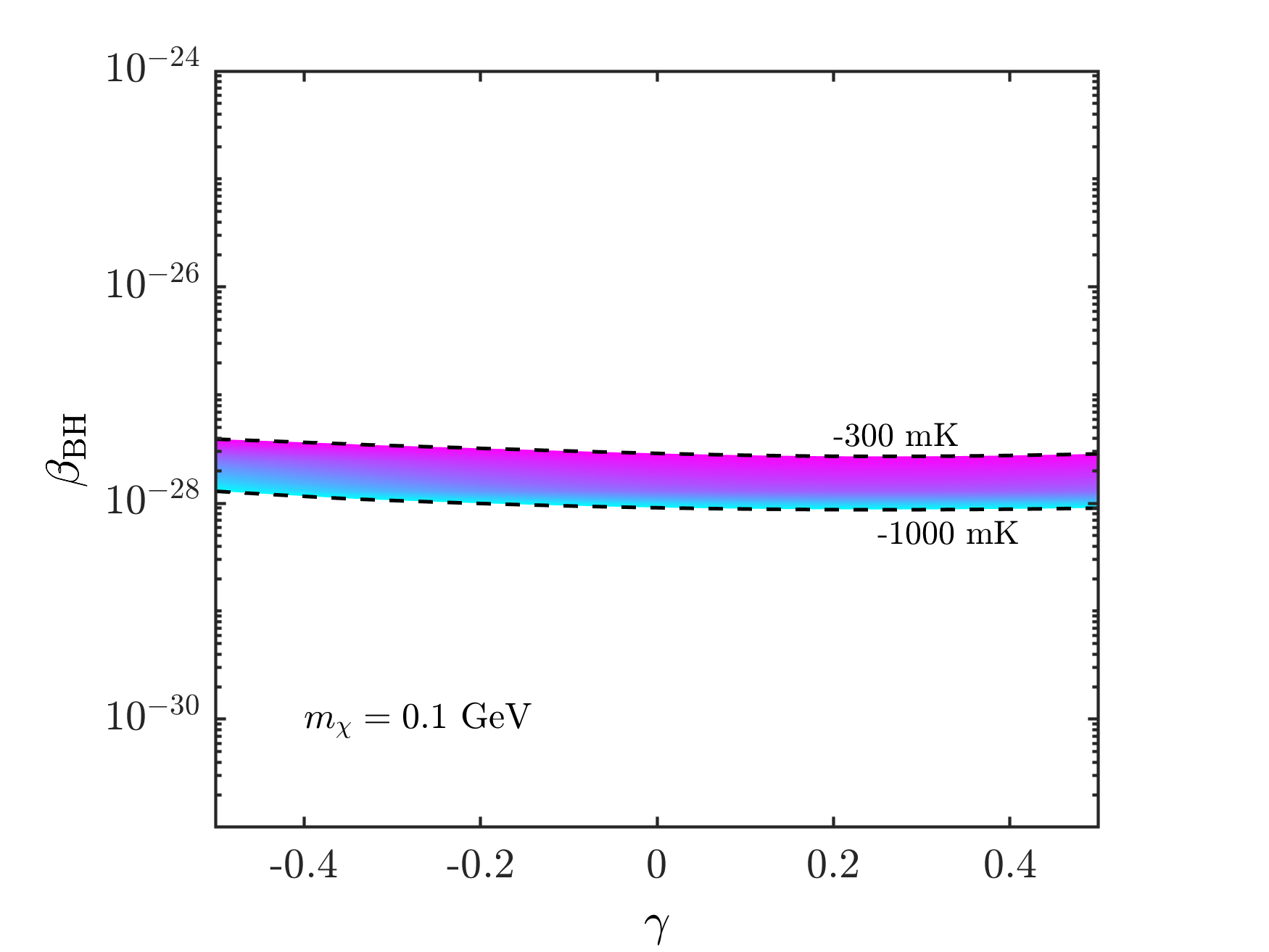}&
		\includegraphics[width=0.5\textwidth]{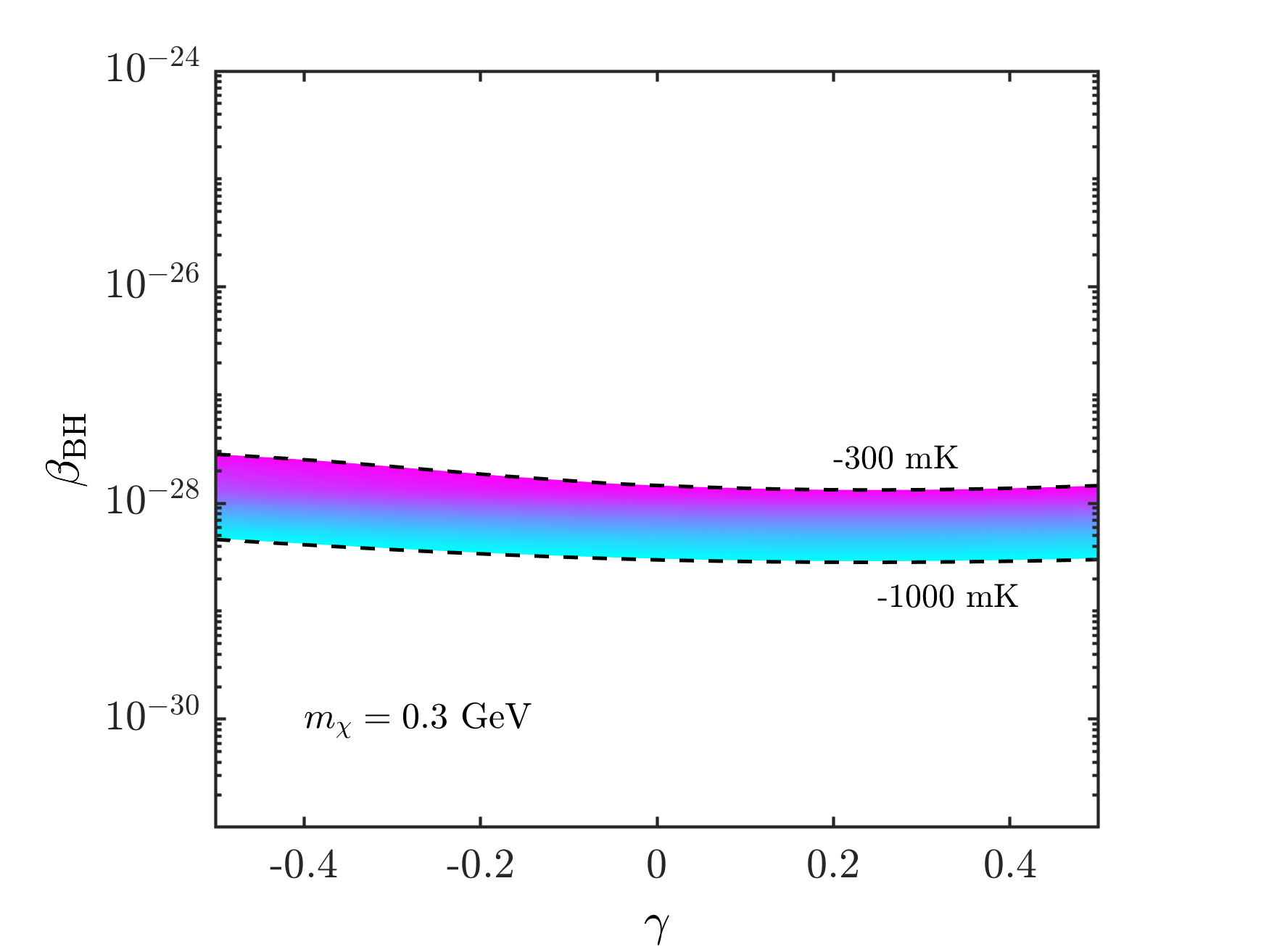}\\
		(a)&(b)\\
		\includegraphics[width=0.5\textwidth]{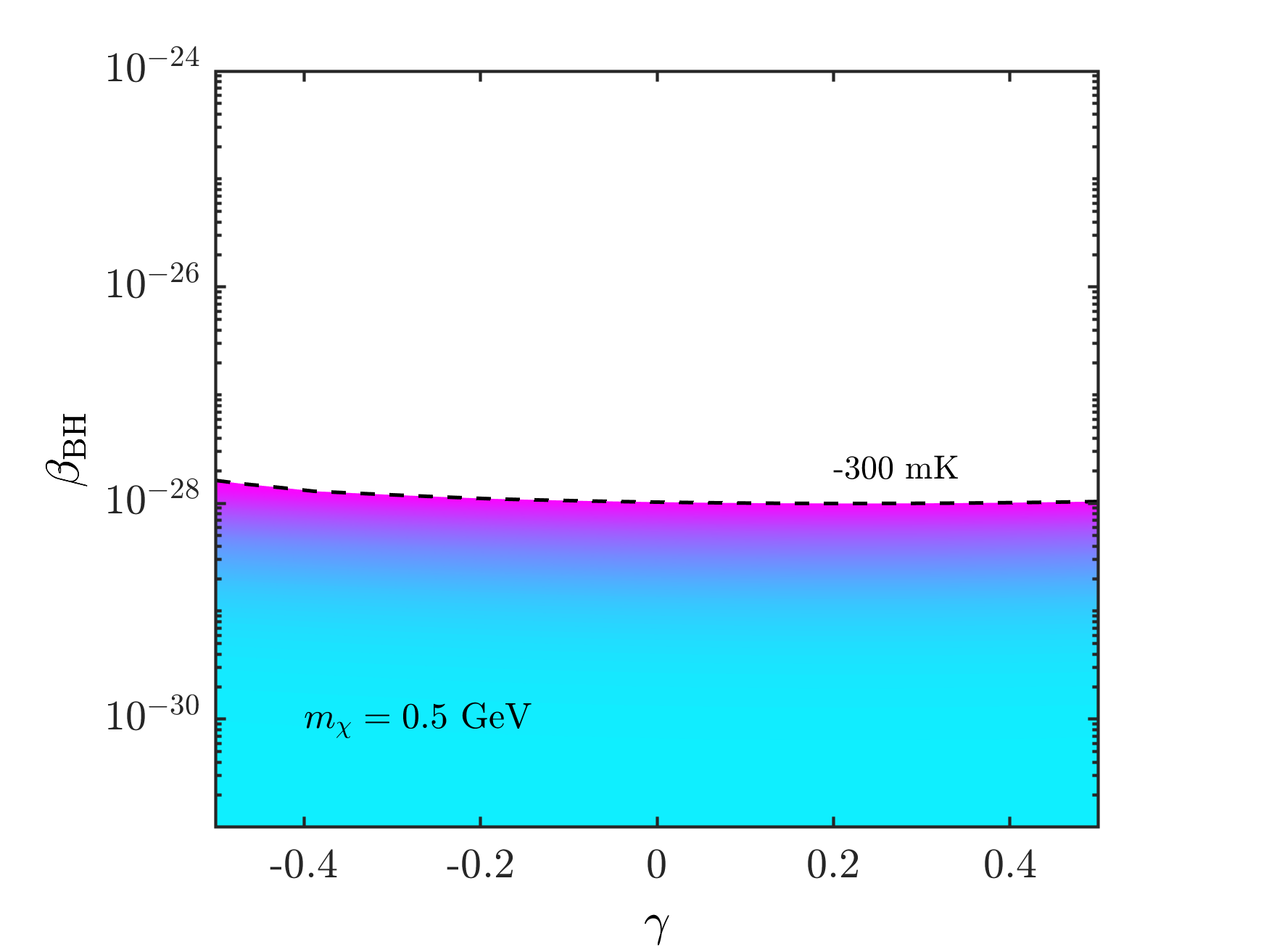}&
		\includegraphics[width=0.5\textwidth]{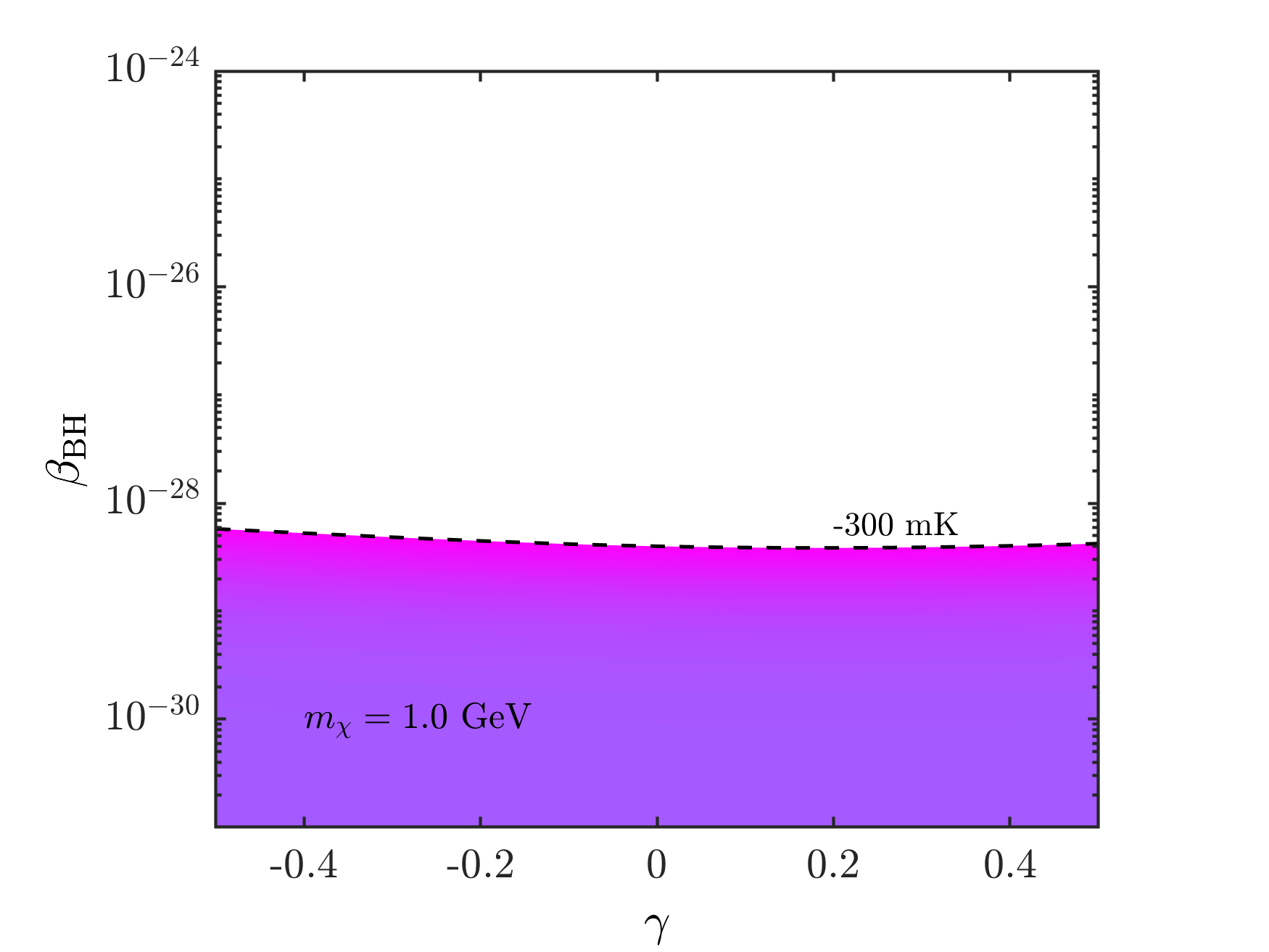}\\
		(c)&(d)\\
	\end{tabular}
	\begin{tabular}{c}
		\includegraphics[trim=0 10 0 310,clip, width=0.7\textwidth]{figure/cbar.pdf}\\
	\end{tabular}
	\caption{\label{fig:pow_beta_gamma} Allowed zones in the $\beta_{\rm BH} - \gamma$ plane for power law mass distribution of PBH, where different values of DM masses are considered ((a) $m_{\chi}=0.1$ GeV, (b) $m_{\chi}=0.3$ GeV, (c) $m_{\chi}=0.5$ GeV and (d) $m_{\chi}=1.0$ GeV). Upper and lower limits of PBH mass distribution are adopted to be $M_{\rm min}=6\times 10^{13}$ g and $M_{\rm max}=10^{15}$ g for these plots.}
\end{figure}

In Fig.~\ref{fig:pow_beta_gamma}, allowed regions (coloured zones) in the $\beta_{\rm BH} - \gamma$ parameter space that satisfy the EDGES results are shown. The calculated value of $T_{21}$ at redshift $z\simeq 17.2$ is denoted by $T_{21}^{z=17.2}$ and is represented by the colour bars in the plots. The allowed zones of the $\beta_{\rm BH} - \gamma$ parameter space are estimated for different chosen values of DM masses (Fig.~\ref{fig:pow_beta_gamma}(a) $m_{\chi}=0.1$ GeV, Fig.~\ref{fig:pow_beta_gamma}(b) $m_{\chi}=0.3$ GeV, Fig.~\ref{fig:pow_beta_gamma}(c) $m_{\chi}=0.5$ GeV and Fig.~\ref{fig:pow_beta_gamma}(d) $m_{\chi}=1.0$ GeV). It can be observed that for $m_\chi$ = 0.1 GeV, the narrowest allowed region of $\beta_{\rm BH} - \gamma$ is obtained among the four cases and the lower limits of the allowed region drop significantly with the increment of DM mass values. This can have similar explanation as in Fig.~\ref{fig:beta_mu} and Fig.~\ref{fig:beta_sigma}. Moreover, it can be observed from Fig.~\ref{fig:pow_beta_gamma} that $\beta_{\rm BH}$ slightly decreases as the power index $\gamma$ increases and hence $\beta_{\rm BH}$ depends on the formation time of the PBH. The value of $\gamma$ is varied from $-\frac{1}{2}$ to $\frac{1}{2}$ in Fig.~\ref{fig:pow_beta_gamma} which corresponds to the variation of the equation of state $\omega$, or the formation epoch of PBHs, from $\frac{1}{3}$ to $-\frac{1}{5}$.
 It can be noted from Fig.~\ref{fig:pow_beta_gamma} that maximum value of $\beta_{\rm BH}$ is obtained for $\gamma=-\frac{1}{2}$ or $\omega=\frac{1}{3}$ which corresponds to the equation of state of the radiation dominated epoch. Therefore, the Fig.~\ref{fig:pow_beta_gamma} estimates that a larger initial mass fraction of PBH (thus larger initial abundance of PBH) is obtained if the formation time of PBH is radiation dominated epoch and the abundance decreases slightly if the PBHs formation take place at later epochs ($\gamma > -\frac{1}{2}$).

\subsection{Probability Distribution of Primordial Black Hole Masses Calculated from Excess 21cm Absorption Results}\label{ssec:edges_dist_asha} 

Two analytical distribution models for PBH mass distribution (lognormal distribution and power law distribution) have been studied in the previous sections (Subsect.~\ref{ssec:lognormal} and \ref{ssec:powerlaw} respectively), which are widely used in literature.

In this section considering different PBH masses  $M_{\rm BH}$ and their monochromatic distributions, we attempted to construct a probability distribution function of PBH masses by utilizing the weightages of different values of 21cm temperature within the extent of $T_{21}^{z=17.2}$ values furnished by EDGES experiment at $z=17.2$.

Different probability distribution functions for different $M_{\rm BH}$ for fixed values of $\beta_{\rm BH}$ can be constructed by utilizing the EDGES 21cm result at reionization epoch. The EDGES result indicates 21cm brightness temperature $T_{21}=-500^{+200}_{-500}$mK. One can construct a distribution probability weightage ($G_{21}$) of the $T_{21}^{z=17.2}$ in the range $-200\leq T_{21}^{z=17.2}\leq -1000$mK obtained from EDGES experiment, where this probability distribution fixed at $T_{21}^{z=17.2}=-500$mK. The temperature $T_{21}^{z=17.2}$ that a PBH with a particular monochromatic mass $M_{\rm BH}$ and a given of $\beta_{\rm BH}$ would result a 21cm brightness temperature $T_{21}^{z=17.2}$ on evaporation. From the probability weightage ($G_{21}$) plot of $T_{21}^{z=17.2}$, the probability weightage for the value of $T_{21}^{z=17.2}$ calculated for a fixed $M_{\rm BH}-\beta_{\rm BH}$ pair can be obtained. Thus an analytical map of weightage probability of $T_{21}^{z=17.2}$ on the distribution of $M_{\rm BH}-\beta_{\rm BH}$ pair can be constructed. Therefore for a fixed value of $\beta_{\rm BH}$, a mapped distribution of $M_{\rm BH}$ (infact $M_{\rm BH}$ vs $G_{21}$) is obtained from $M_{\rm BH}$ - $\beta_{\rm BH}$ - $G_{21}$ constructed using the procedure described here.

The EDGES result for $T_{21}$ and its allowed extent are used to obtain the final form of such a proposed distribution. The distribution and analysis described in Sect.~\ref{sect. PBH} -~\ref{sect. T21} are utilized to propose and explore this analytical distribution by PBHs. As has been discussed earlier, the EDGES observation has reported the brightness temperature at the redshift $z\sim 17.2$ is $T_{21}=-500^{+200}_{-500}$ mK with 99$\%$ confidence level. Consequently, the probabilities of different brightness temperature within the range $-1000$ mK $\leq T_{21}\leq$ $-300$ mK can be computed at that epoch.

Firstly, a basic skew normal distribution for $T_{21}^{z=17.2}$ (see Eq.~\ref{eq:snd}) is fitted in such a way that, the peak of the distribution lie at $-500$ mK and $99\%$ of this distribution lies within the range $-1000$ mK $\leq T_{21}^{z=17.2}\leq$ $-300$ mK while the individual probabilities at  $T_{21}=-300$ mK and at $-1000$ mK are equal.

	The fitted skew normal distribution (SND) function given by,
	\begin{equation}
		G_{21}=\left(1+{\rm erf} \left(\dfrac{\alpha_1 (T_{21}^{z=17.2}-\mu_1)}{\sqrt{2}\sigma_1}\right)\right)\exp \left(-\dfrac{(T_{21}^{z=17.2}-\mu_1)^2}{2 \sigma_1^2}\right),
		\label{eq:snd}
	\end{equation}
	where the fitted parameters are $\mu_1=-405.708$, $\sigma_1=223.35$ and $\alpha_1=-3.90017$ (in the above equation $T_{21}^{z=17.2}$  is $T_{21}$ at redshift $\sim 17.2$). It is to be mentioned that, the above expression (Eq.~\ref{eq:snd}) describes only the probabilities of brightness temperature $T_{21}^{z=17.2}$ (i.e., $T_{21}$ at $z \simeq 17.2$). The relation between PBH mass and $T_{21}^{z=17.2}$ is to be calculated numerically for different chosen DM mass ($m_{\chi}$) and PBH initial mass fractions ($\beta_{\rm BH}$). 
	\begin{figure}
		\centering
		\includegraphics[width=0.6\linewidth]{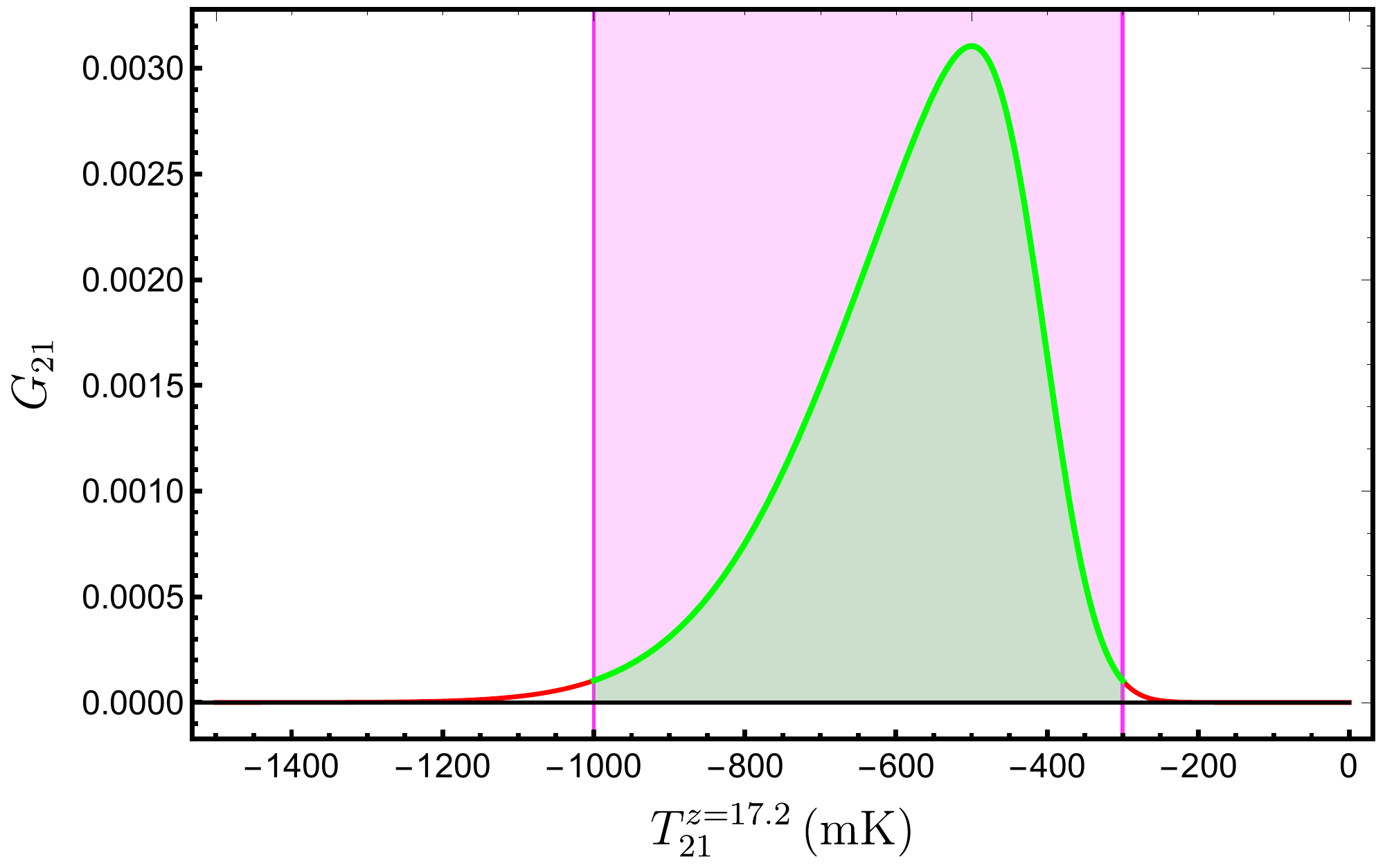}
		\caption{\label{fig:g21_t21} The fitted function (Eq.~\ref{eq:snd}) of $G_{21}$ using EDGES observational result. The green solid line denotes $G_{21}$ as function of $T_{21}^{z=17.2}$. The pink region represents the uncertainty of the EDGES result (i.e. $-1000 \sim -300$ mK), while the solid green region occupies the 99\% confidence level.}
	\end{figure}
	Eq.~\ref{eq:snd} is graphically described in Fig.~\ref{fig:g21_t21}. In this plot the 99\% C.L. region of the probability distribution function $G_{21}$ for $T_{21}^{z=17.2}$ in the range -1000 mK $\leqslant T_{21}^{z=17.2} \leqslant$ -300 mK are shown by the green region. It is to be noted that $G_{21}$ at both the boundaries (i.e. $-300$ mK and $-1000$ mK) are kept equal. 
	
In order to obtain analytical form of the probability distribution of PBH masses, the values of $T_{21}^{z=17.2}$ are numerically calculated for different PBH masses ($M_{\rm BH}$) by solving coupled equations (Eqs.~\ref{mdecay},~\ref{T_chi},~\ref{T_b},~\ref{xe} and \ref{V_chib}) for each possible combinations of PBH parameters by considering monochromatic distribution of PBHs. In this case we adopt the procedure and corresponding equations as introduced in Ref.~\cite{Halder:2021rbq}. In the entire calculation we choose the DM mass $m_{\chi}=0.5$ GeV and $\sigma_{41}=1$ as benchmark values.
	
	\begin{figure}
		\centering
		\begin{tabular}{ccc}
			\includegraphics[trim=10 0 35 0, clip, width=0.42\textwidth]{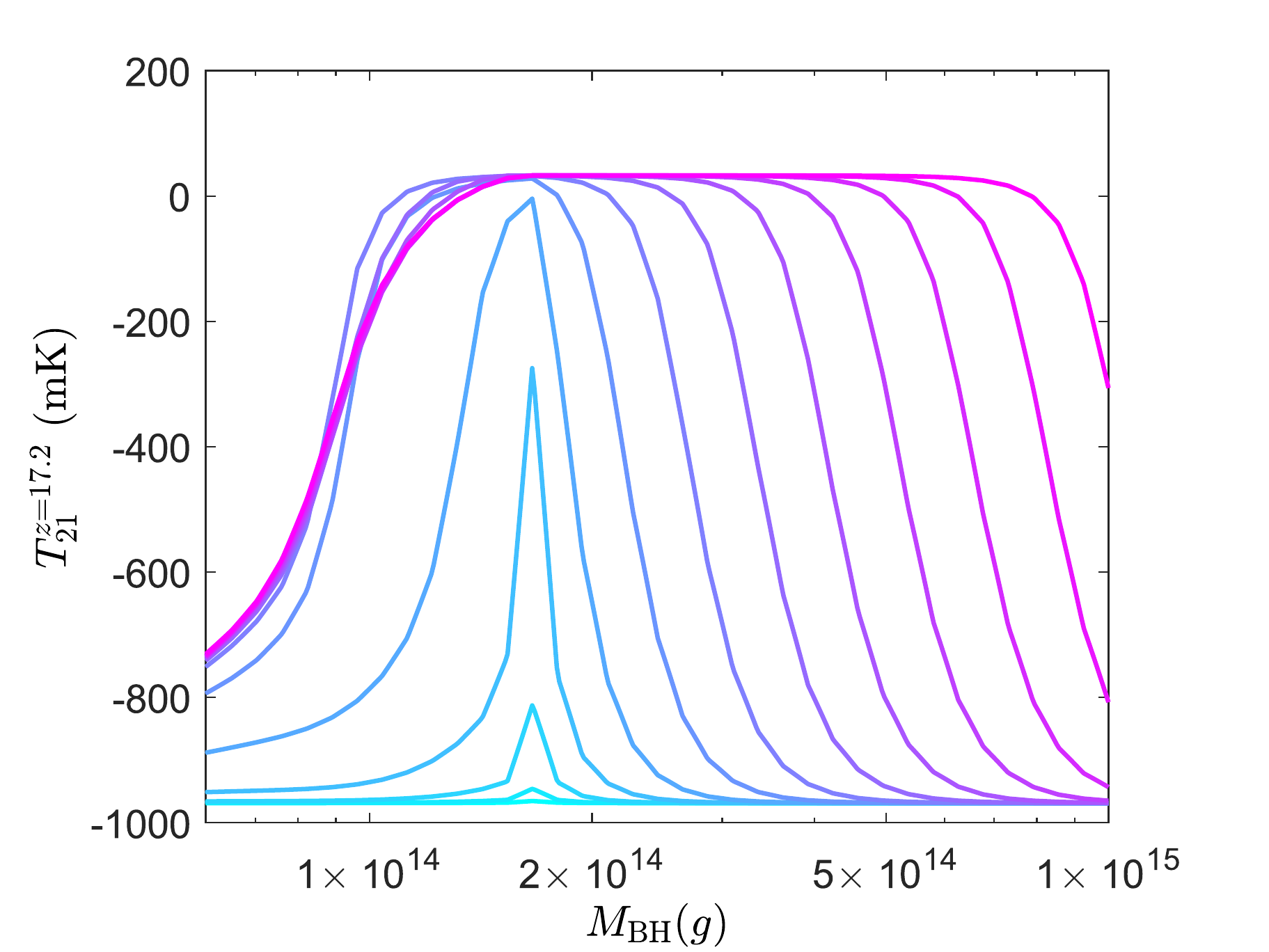}&
			\includegraphics[trim=5 0 35 0, clip, width=0.42\textwidth]{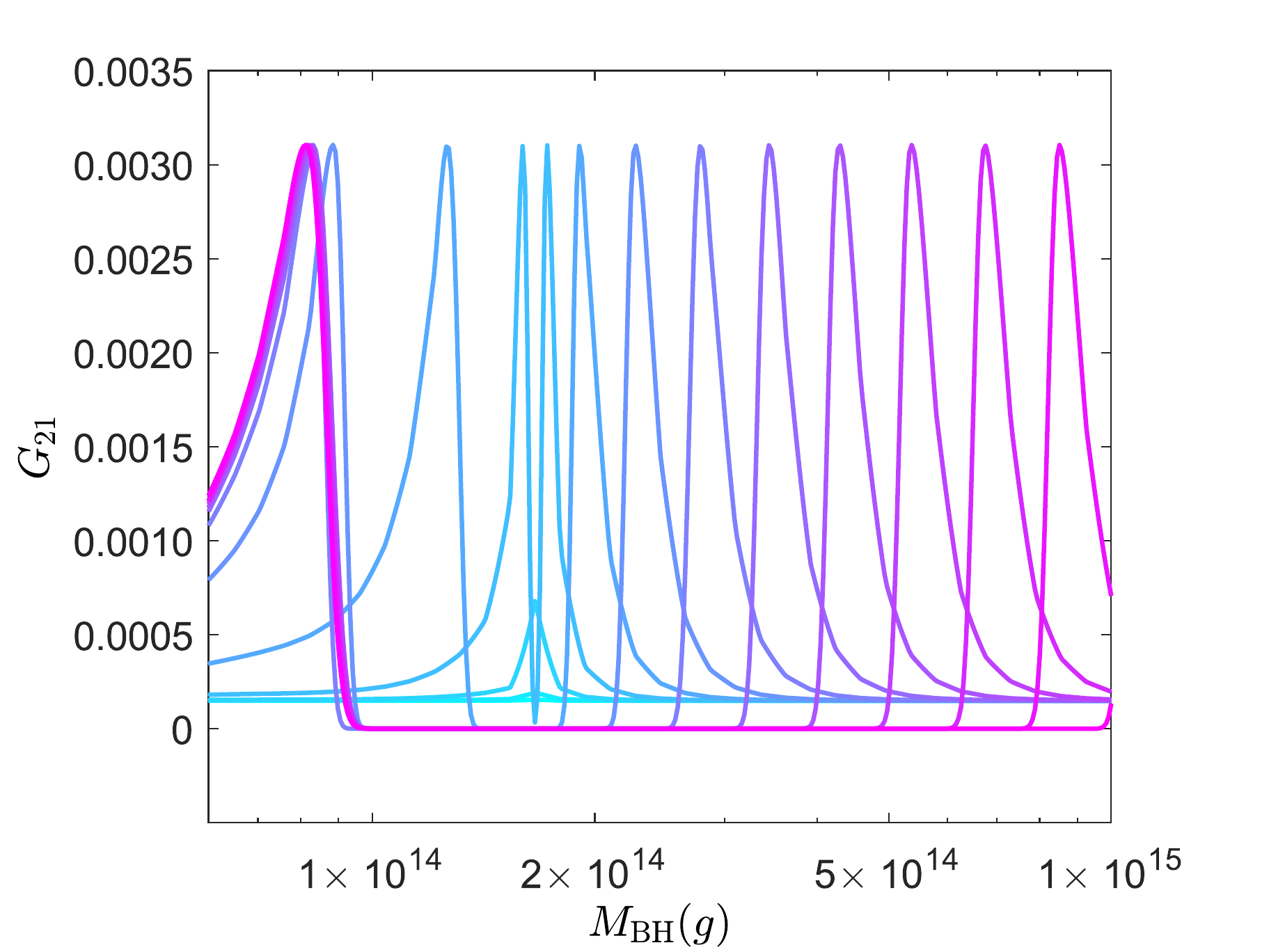}&
			\includegraphics[height=5cm,width=0.09\linewidth,trim={260 70 190 40},clip]{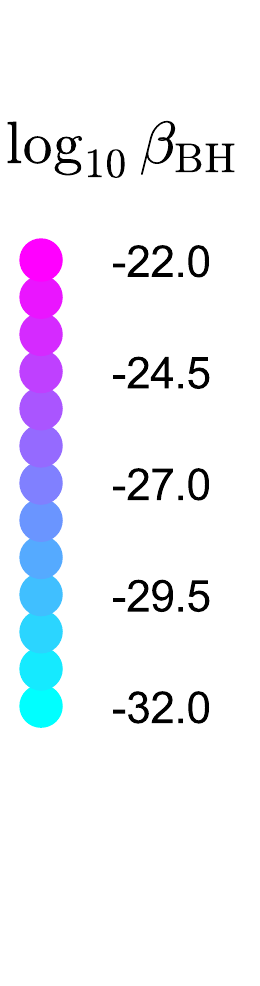}\\
			(a)&(b)&\\
			\includegraphics[trim=10 0 35 0, clip, width=0.42\linewidth]{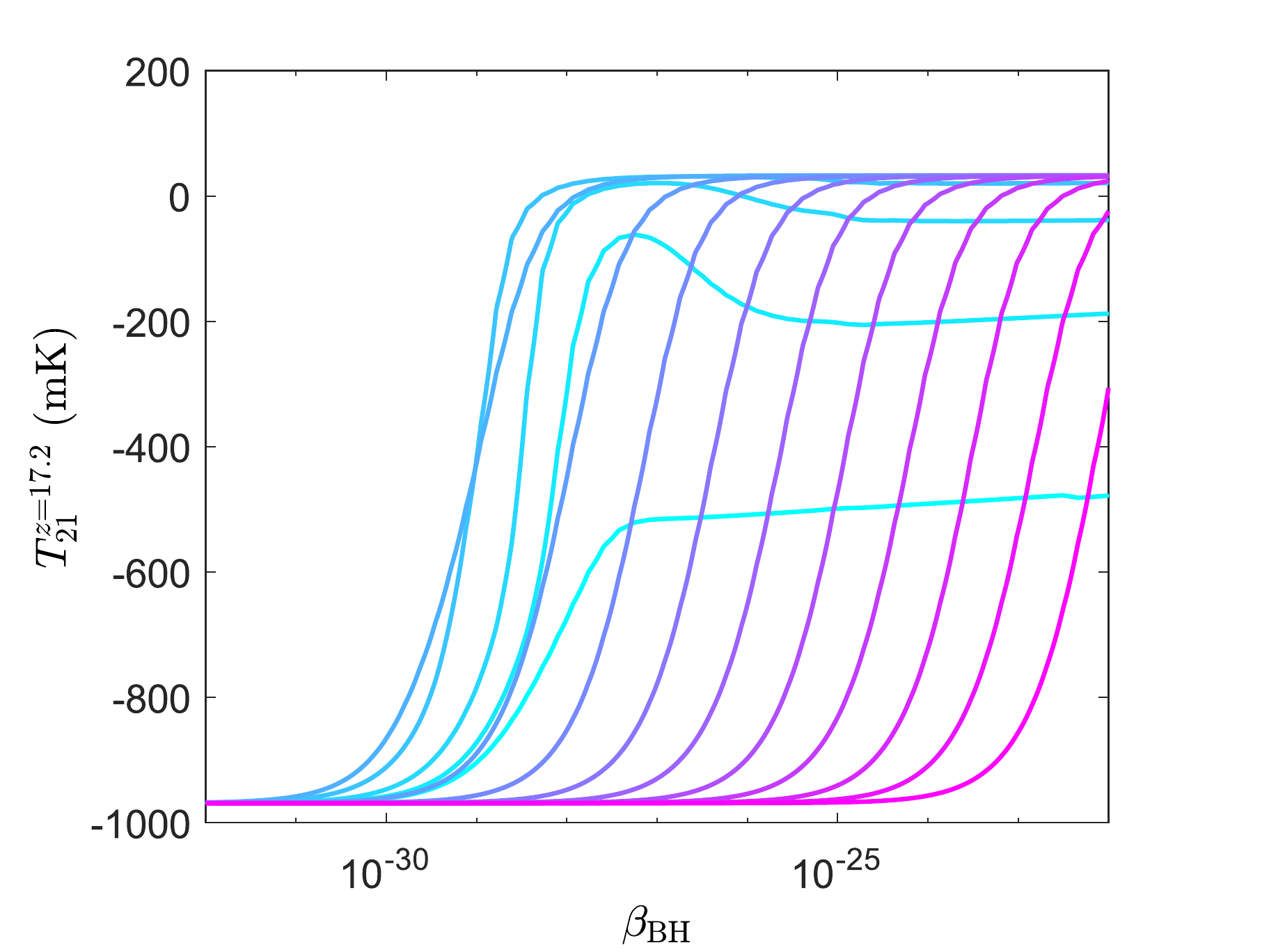}&
			\includegraphics[trim=5 0 35 0, clip, width=0.42\linewidth]{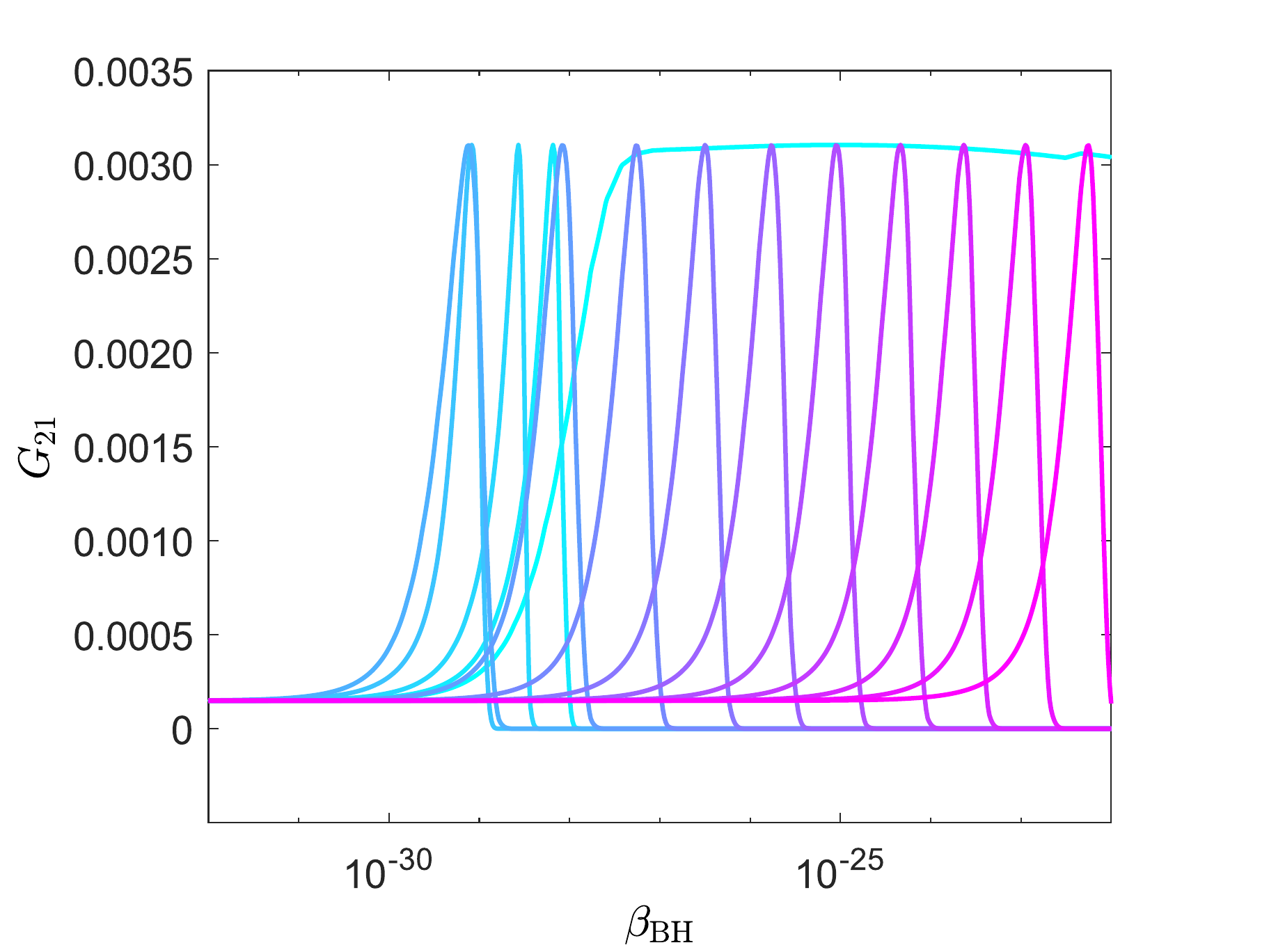}&
			\includegraphics[height=5cm,width=0.12\linewidth,trim={260 60 150 40},clip]{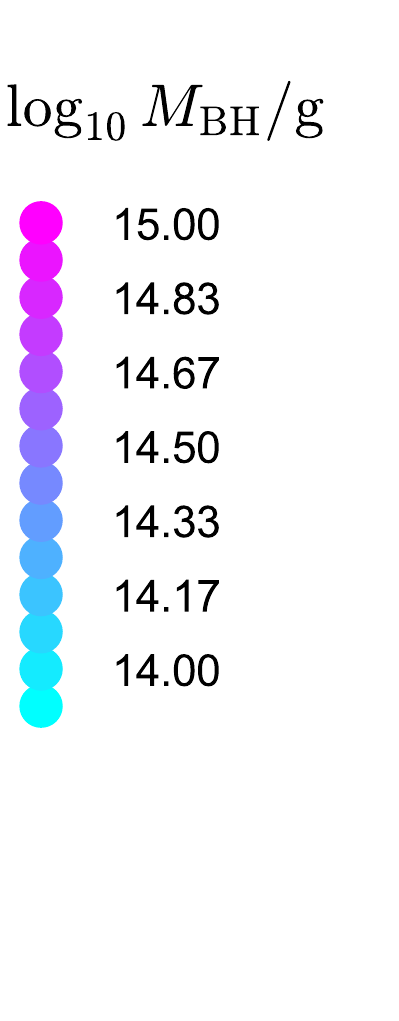}\\
			(c)&(d)&\\
		\end{tabular}
		\caption{\label{GT-beta} Variations of $T_{21}^{z=17.2}$ (left column) and corresponding $G_{21}$ (right column) with $M_{\rm BH}$ for different values of $\beta_{\rm BH}$ are shown in the plots of upper row (Fig.~\ref{GT-beta}(a) and \ref{GT-beta}(b)), while the similar variations with $\beta_{\rm BH}$ for different are plotted in the graphs of lower row (Fig.~\ref{GT-beta}(c) and \ref{GT-beta}(d)). Different chosen values of $M_{\rm BH}$ ($\beta_{\rm BH}$) are represented by different colours, as mentioned in the colour bar at the end of upper (lower) row.}
	\end{figure}

	In Fig.~\ref{GT-beta} the variations of $T_{21}^{z=17.2}$ and corresponding $G_{21}$ for different masses $M_{\rm BH}$ of PBHs are graphically described. Fig.~\ref{GT-beta}(a) addresses the variation of $T_{21}^{z=17.2}$ vs PBH mass $M_{\rm BH}$ for different initial mass fractions of primordial black holes $\beta_{\rm BH}$. Different values of $\beta_{\rm BH}$ are represented by the lines of different colours, where the corresponding values of $\beta_{\rm BH}$ are shown by the colour bar, given at the right of top panel of Fig.~\ref{GT-beta}. The corresponding $G_{21}$ vs $M_{\rm BH}$ plots are furnished in Fig.~\ref{GT-beta}(b). Figs.~\ref{GT-beta}(a) and \ref{GT-beta}(b) are similar except that in Fig. \ref{GT-beta}(b) the variaitons of $G_{21}$ with $M_{\rm BH}$ are shown. The plots if Fig. \ref{GT-beta}(b) are generated by identifying $T_{21}^{z=17.2}$ value corresponding to $G_{21}$ value (from Fig. \ref{fig:g21_t21} or Eq.~\ref{eq:snd}) and then following the same procedure to obtain Fig. \ref{GT-beta}(a). Similar representations of $T_{21}^{z=17.2}$ vs $\beta_{\rm BH}$ and corresponding $G_{21}$ vs $\beta_{\rm BH}$ for different chosen values of $M_{BH}$ are shown in \ref{GT-beta}(c) and \ref{GT-beta}(d) respectively. From Fig.~\ref{GT-beta}(a) (and also from Fig.~\ref{GT-beta}(b)), it can be noticed that, for lower values of $\beta_{\rm BH}$s all the $T_{21}^{z=17.2}$ vs $M_{\rm BH}$ graphs suffer certain discontinuities near $M_{\rm BH}\approxeq 1.7 \times 10^{14}$ g. Such nature arises as the PBHs of mass $M_{\rm BH}\approxeq 1.7 \times 10^{14}$ g  had completely evaporated at $z \sim 17$. Similar features are also obtained from Fig.~\ref{GT-beta}(c) and \ref{GT-beta}(d) for lower values of $\beta_{\rm BH}$.
	
	\begin{figure}
		\centering
		\begin{tabular}{cc}
			\includegraphics[width=0.5\linewidth]{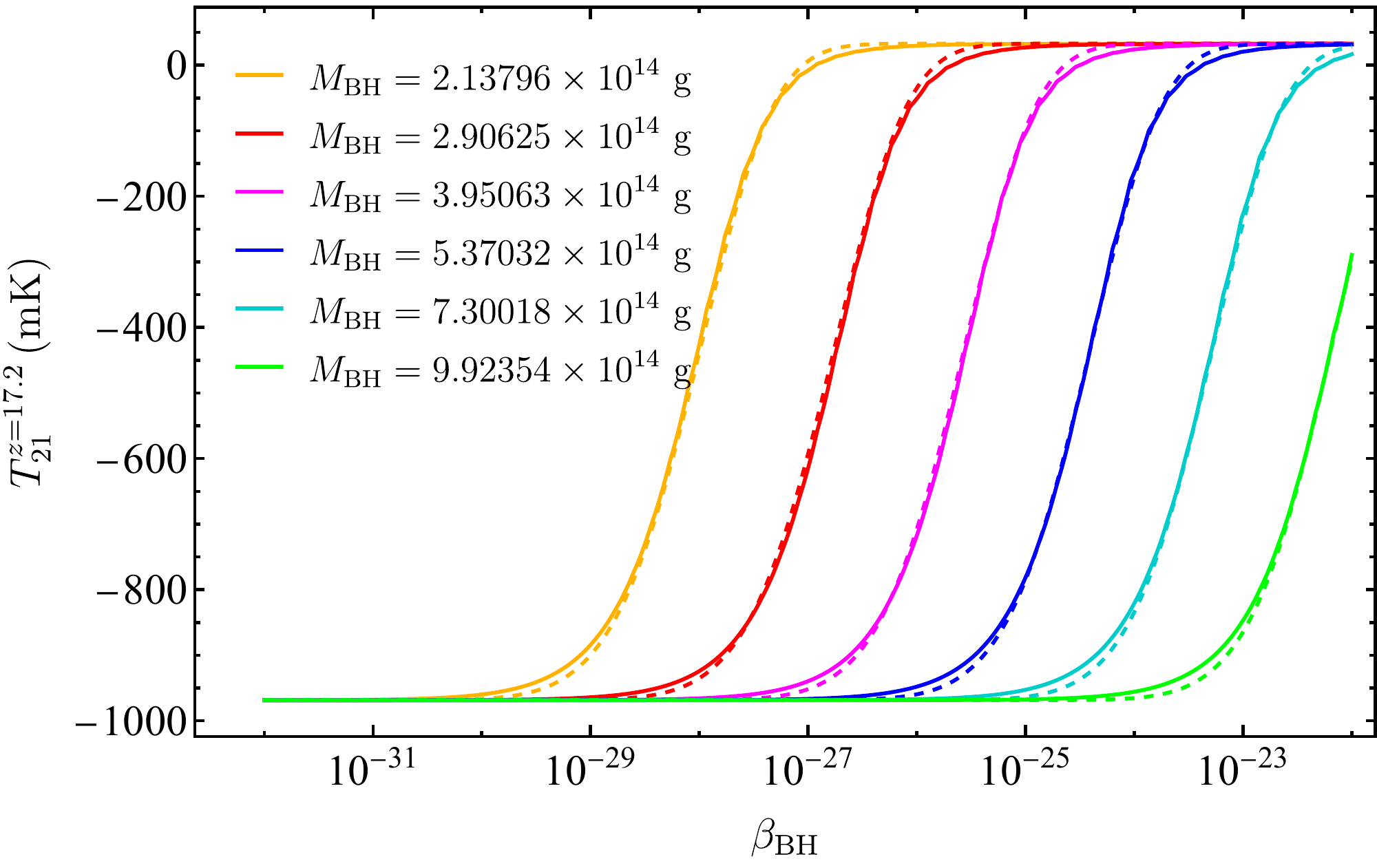}&
			\includegraphics[width=0.5\linewidth]{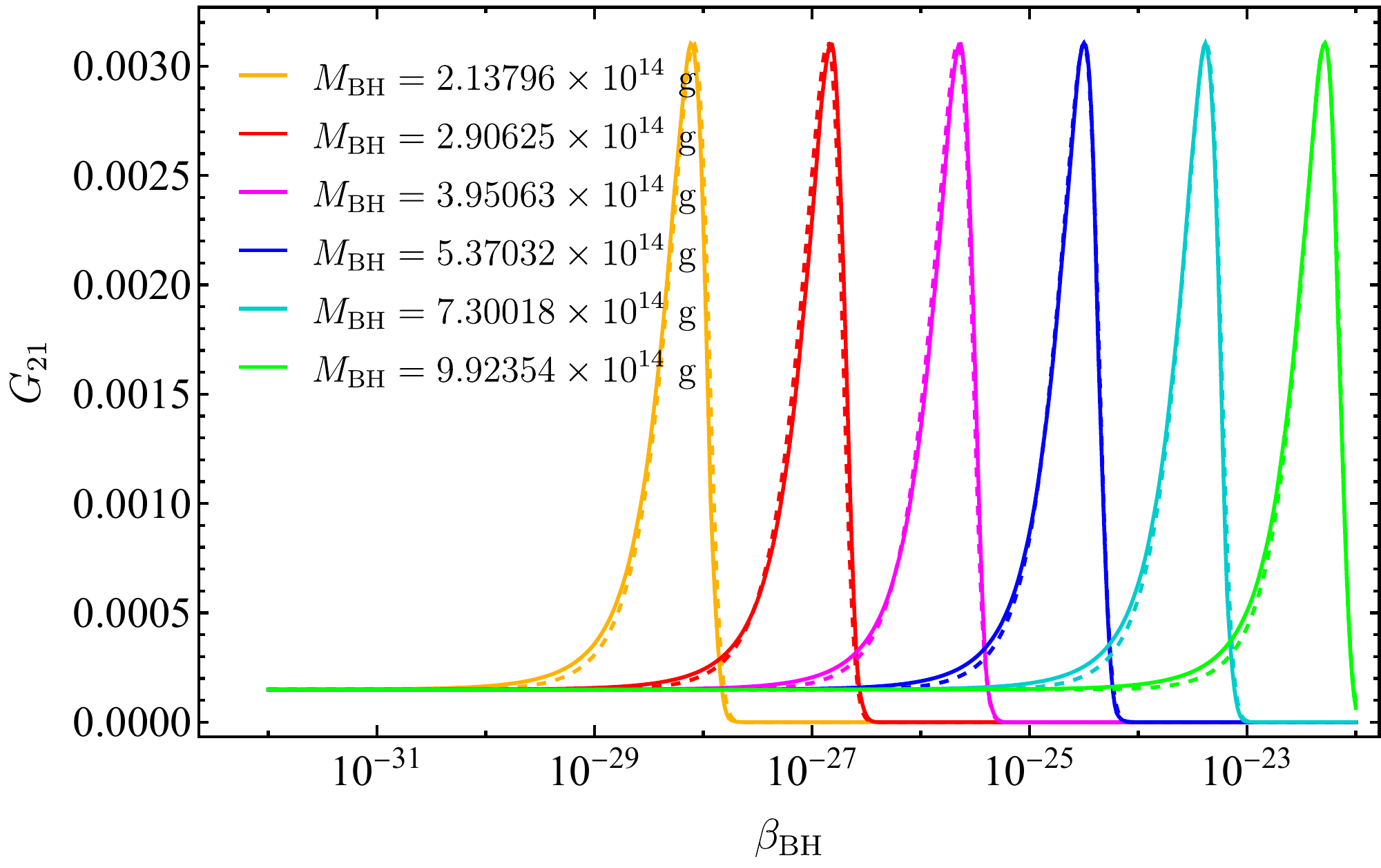}\\
			(a)&(b)\\
		\end{tabular}
		\caption{Dotted lines are showing values obtained with Eq.~\ref{eq:snd} and  \ref{eq:fit_t21} and solid lines are representing values obtained by solving suitable equations of sections \ref{sect. PBH} - \ref{sect. T21}. \label{GT_fitted}}
	\end{figure}

	In the present analysis our attempt is to find an approximately fitted analytical form of $G_{21}$ (for $M_{\rm BH}\gtrapprox 2 \times 10^{14}$ g) as a  function of PBH mass $M_{\rm BH}$ and initial PBH mass fraction $\beta_{\rm BH}$. In this regard we first propose a form for $T_{21}^{z=17.2}$ as,
	\begin{equation}
		T_{21}^{z=17.2}(M_{\rm BH}, \beta_{\rm BH})=(a_1 + a_2)\times {\rm CDF_{SQN}}(M_{\rm BH},\beta_{\rm BH},\sigma_f,\alpha_f)-a_2,
	\label{eq:fit_t21}
	\end{equation}
	where $a_1$, $a_2$, $\mu_f$, $\sigma_f$, $\alpha_f$ are parameters obtained by fitting this equation with the results obtained in Fig. \ref{GT-beta}. The cumulative distribution ${\rm CDF_{SQN}}$ (in Eq.~\ref{eq:fit_t21}) is given by
	\begin{equation}
		{\rm CDF_{SQN}}(M_{\rm BH},\beta_{\rm BH},\sigma_f,\alpha_f)=\dfrac{1}{2}{\rm erfc}\left(\dfrac{\mu_f(M_{\rm BH})-\log_{10}\beta_{\rm BH}}{\sqrt{2} \alpha_f}\right)-2~{\rm OwenT}\left(\dfrac{\log_{10}\beta_{\rm BH}-\mu_f(M_{\rm BH})}{\sigma_f},\alpha_f \right).
	\label{eq:fit_cdf}
	\end{equation}
	In the above expression, ${\rm erfc}(x)$ and ${\rm OwenT}\left(i,j \right)$ are the complementary error function and the Owen function \cite{owen_f} respectively, given by
	\begin{equation}
		{\rm erfc}(x)=1-{\rm erf}(x)
		\label{eq:erfcf}
	\end{equation}
	\begin{equation}
		{\rm OwenT}(x,a)=\dfrac{1}{2 \pi} \displaystyle\int_0^a \exp\left(\dfrac{-x^2 (1+t^2)/2}{1+t^2}\right) dt\,\,.
		\label{eq:owentf}
	\end{equation}
	The parameters are obtained as $a_1 = 32.83$, $a_2=968.576$ and $\alpha_f=-1.5$ We also note that values of $\sigma_f$ and $\mu_f$ vary with the PBH mass $M_{\rm BH}$ while the other parameter $\alpha_f$ remains almost unchanged. The variations of $\sigma_f$ and $\mu_f$ with $M_{\rm BH}$ are found to be approximated as,
	\begin{eqnarray}
		\sigma_f(M_{\rm BH})&\approxeq& 1.66 - 0.026 \log_e (M_{\rm BH}/{\rm g}), \\
		\mu_f(M_{\rm BH})&\approxeq& 13.0584 - 1100.38 \left(M_{\rm BH}/{\rm g}\right)^{-0.1}. \label{eq:mu_f}
	\end{eqnarray}
It should be mentioned here that, the form of the fitted function in Eq.~\ref{eq:fit_t21} is obtained by trial and $\chi^2$-fitting. It can now be seen that, replacing $T_{21}^{z=17.2}$ in Eq.~\ref{eq:snd} with the expression of $T_{21}^{z=17.2}$ obtained in Eq.~\ref{eq:fit_t21}, a form for a distribution of PBH masses follows (i.e. $G_{21}(M_{\rm BH},\beta_{\rm BH}) = f(M_{\rm BH},\beta_{\rm BH})$). But it is to be noted here that numerical values of the parameters of this distribution function would change for different chosen DM mass $m_\chi$ In figure 7(a,b), we compare the results obtained from the analytical form for  $G_{21}$ and $T^{z=17.2}_{21}$ (eqs 5.1-5.6; proposed in this work) with those obtained from the simulations.
	
\begin{figure}
		\centering
		\includegraphics[width=0.6\linewidth]{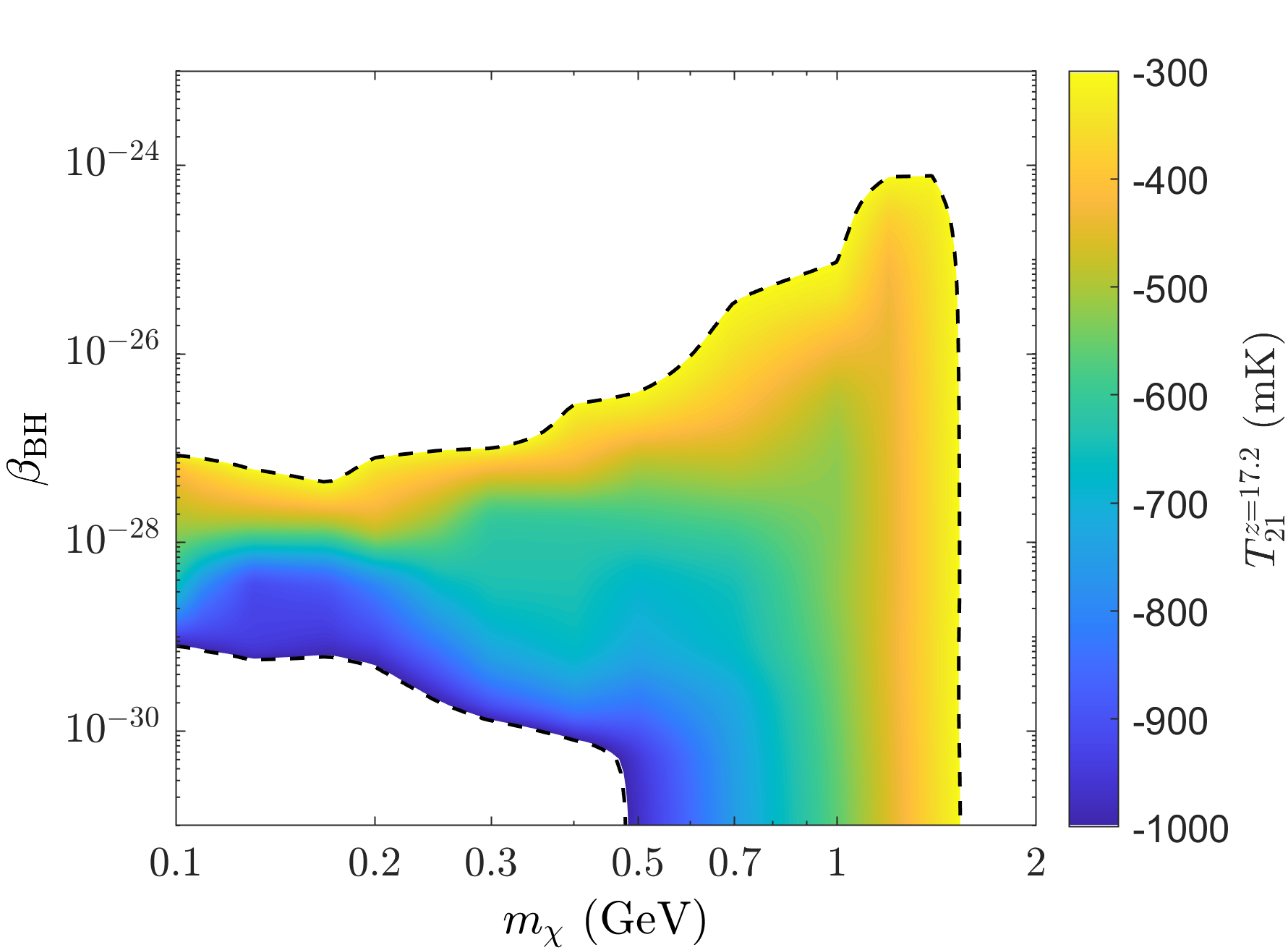}
		\caption{\label{fig:beta_mchi} The allowed region in the $\beta_{\rm BH}$ - $m_{\chi}$ space that satisfies the 21cm brightness temperature limit $\left(-500^{+200}_{-500}\right)$ (EDGES result). The black dashed lines represent the upper and lower limits for the same, while the different colours of the contour plot denote different values of $T_{21}^{z=17.2}$ in mK.}
	\end{figure}

Since the evolution of the spin temperature and consequently the brightness temperature of 21
cm line also depend on DM mass $m_{\chi}$, we repeat the entire calculation with different chosen values of Dark Matter mass $m_{\chi}$ (in the previous case we use fixed value of Dark Matter particle $m_{\chi}=0.5$ GeV). However, the values of fitted parameters $a_1$, $a_2$, $\mu_f$, $\sigma_f$ and $\alpha_f$ may be modified with different choices of $m_{\chi}$ but the functional form of $G_{21}$ remains same. Instead of showing the variation of those parameters with $m_{\chi}$, in Fig.~\ref{fig:beta_mchi} the allowed region in the $\beta_{\rm BH}$ - $m_{\chi}$ parameter space is plotted.
Fig.~\ref{fig:beta_mchi} is plotted in the following way. In the discussion above, it is clear that for a particular value of $\beta_{\rm BH}$ a probability weightage distribution function is obtained. Now using this $\beta_{\rm BH}$ value and the corresponding probability distribution function ($G_{21}$) of $M_{\rm BH}$, the value of $T_{21}^{z=17.2}$ is computed (solving the coupled equations given in sections \ref{sect. PBH}-\ref{sect. T21}) for different values of dark matter mass $m_{\chi}$ within a chosen range of $m_{\chi}$ (0.1 GeV$\sim$ 2 GeV). The values of $\beta_{\rm BH}$ and $m_{\chi}$ which yield $T_{21}^{z=17.2}$ within the allowed range of EDGES result are then plotted. This process is repeated for different values of $\beta_{\rm BH}$ and the corresponding distribution function for $M_{\rm BH}$. The result is then plotted in Fig.~\ref{fig:beta_mchi} and the allowed region is furnished as shown. Note that, in generating Fig.~\ref{fig:beta_mchi} the mass distribution of PBH masses is considered as mentioned in Eq.~\ref{eq:snd}-\ref{eq:mu_f} with PBH mass range (distribution function) $6\times 10^{13}$ g $\leq M_{\rm BH}\leq10^{15}$ g for different fixed values of $\beta_{\rm BH}$.

\begin{figure}
	\centering
	\begin{tabular}{cc}
	\includegraphics[trim=0 35 45 0, clip, width=0.5\linewidth]{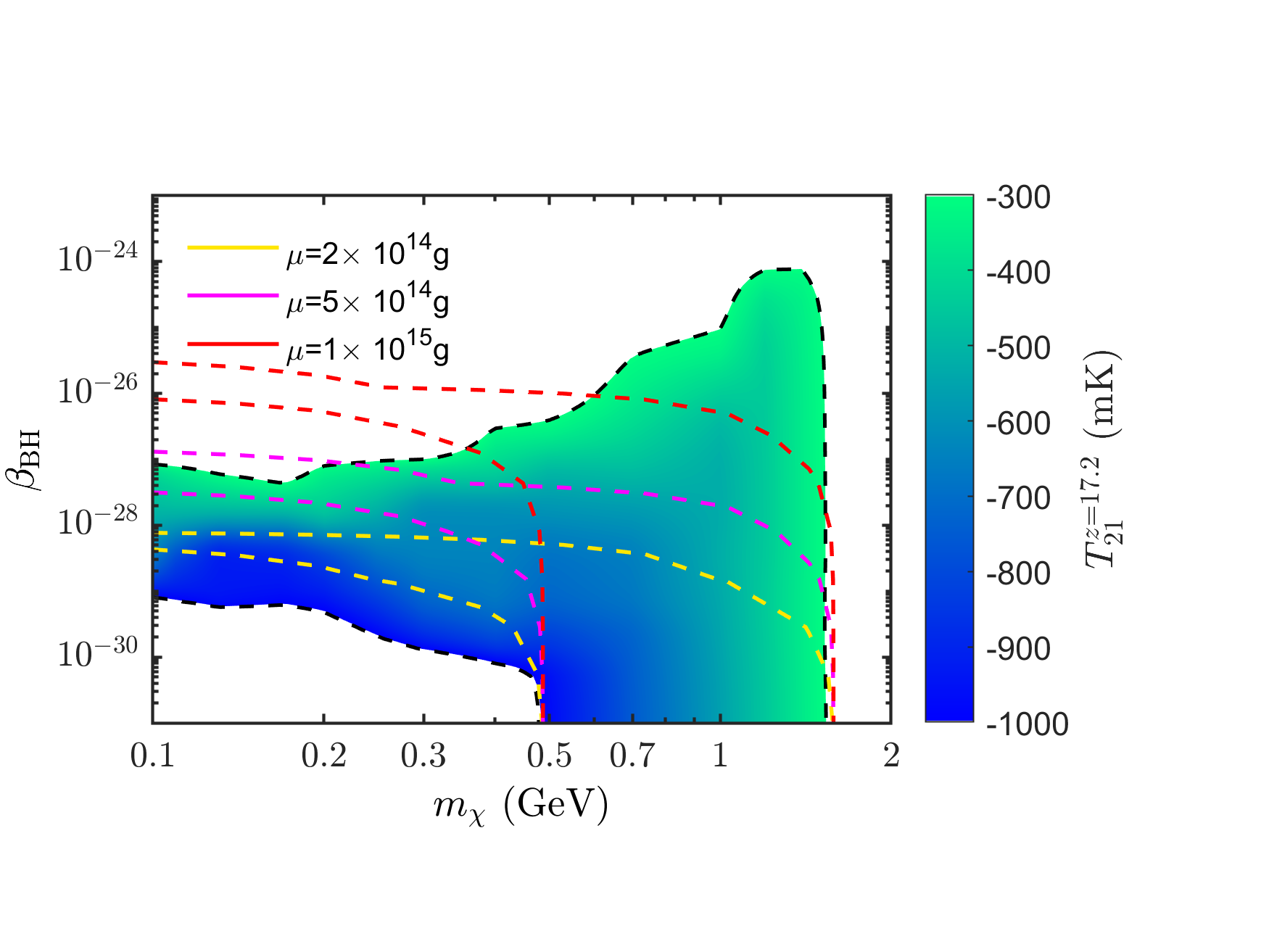}&
	\includegraphics[trim=0 35 45 0, clip, width=0.5\linewidth]{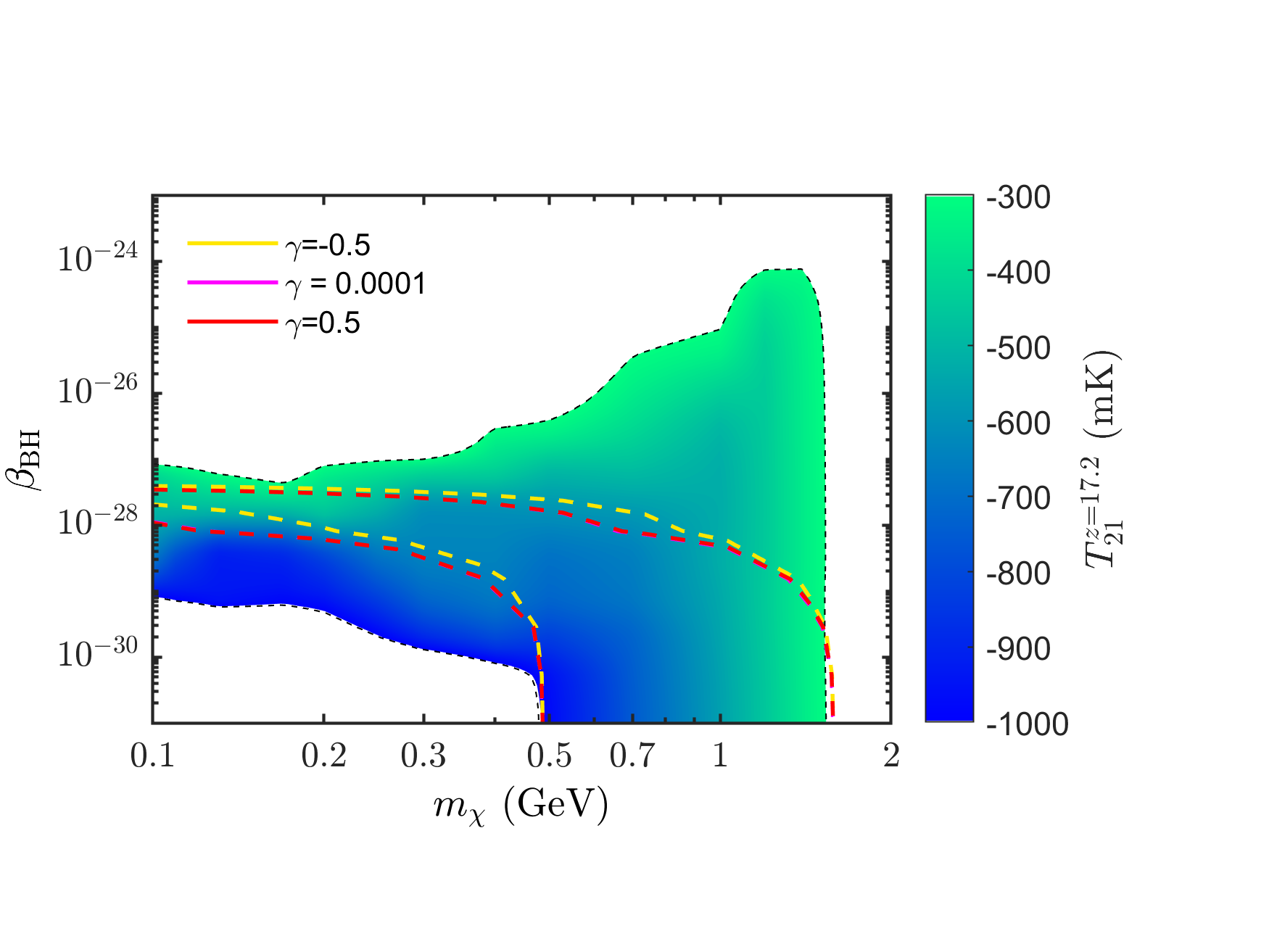}\\
	(a)&(b)\\
	\end{tabular}
	\caption{\label{fig:beta_mx} (a) Allowed region in the $\beta_{\rm BH}-m_{\chi}$ plane by using the EDGES result. The coloured dashed lines of same colour denote the upper and lower bounds for different  chosen values of $\mu$. (b) Allowed region in the $\beta_{\rm BH}-m_{\chi}$ plane by using the EDGES result. The coloured dashed lines of same colour denote the upper and lower bounds for different  chosen values of $\gamma$. Note that, Fig.~\ref{fig:beta_mx}(a) is for the lognormal distribution whereas Fig.~\ref{fig:beta_mx}(b) is related to power law distribution.}
\end{figure}

We also compare the allowed region obtained in Fig.~\ref{fig:beta_mchi} with similar allowed region obtained for other analytical mass distribution functions of PBHs namely lognormal distribution and power law distribution. In case of lognormal distribution of PBH masses the parameters are $\mu$ and $\sigma$ (mean and standard deviation). As discussed earlier the value of $\sigma$ is fixed at a $\sigma=0.5$ lognormal distribution in this work. The coupled differential equations (formalism described in Sect.~\ref{sect. PBH} -~\ref{sect. T21}) are then solved with lognormal distribution of PBHs for several values of $\beta_{\rm BH}$ and $m_{\chi}$, after adopting a suitable fixed value of $\mu$ and the 21cm brightness temperature $T_{21}^{z=17.2}$ is obtained. Then line in the $\beta_{\rm BH}$-$m_{\chi}$ plane that denotes the upper bound of $T_{21}^{z=17.2}$ value ($-300$ mK) is drown in Fig.~\ref{fig:beta_mx}(a) for a particular value of $\mu$ and in the same way line that denotes the lower limit of $T_{21}^{z=17.2}$ temperature ($-1000$ mK) is also shown in the same figure (Fig.~\ref{fig:beta_mx}(a)). The region between these two lines (for a fixed $\mu$ value) is then a allowed region in $\beta_{\rm BH}$-$m_{\chi}$ plane that satisfy EDGES result (in case of lognormal distribution with a given $\mu$ and $\sigma$ value). This process is carried out for three fixed $\mu$ values namely $\mu=2\times 10^{14}$ g, $5\times 10^{14}$ g and $1\times 10^{15}$ g and three allowed regions are obtained which are shown in Fig.~\ref{fig:beta_mx}(a). These regions (bounded by a pair of lines has discussed above) are superimposed in Fig.~\ref{fig:beta_mx}(a) over the region obtained in Fig.~\ref{fig:beta_mchi}, for comparison. Similar plots are drawn and superimposed on the allowed region obtained from Fig.~\ref{fig:beta_mchi} for the case of power law distribution of PBHs and these are shown in Fig.~\ref{fig:beta_mx}(b). Similar procedures for the computations of $T_{21}^{z=17.2}$ are repeated for power law distribution and the upper and lower limits in $\beta_{\rm BH}$-$m_{\chi}$ plane that satisfy EDGES results are shown in Fig.~\ref{fig:beta_mx}(b) and these are superimposed on the allowed region obtained in Fig.~\ref{fig:beta_mchi} for comparison. Note that, for power law distribution there is only one parameter namely the power law index $\gamma$. In Fig.~\ref{fig:beta_mx}(b), the allowed bounded regions (the region between the upper and lower limits of $T_{21}^{z=17.2}$ ($-300$ mK and $-1000$ mK respectively) given by the EDGES experiment) are shown for three fixed value of $\gamma$ namely $\gamma=-0.5$, $0.0001$ and $0.5$ with $M_{\rm min}=6\times 10^{13}$ g and $M_{\rm max}=10^{15}$ g. It is observed from Fig.~\ref{fig:beta_mx}(b) that when $\gamma \geq 0$, the allowed regions appear to coincide.

From Fig.~\ref{fig:beta_mx}(a) and its comparison with Fig.~\ref{fig:beta_mchi} one can see that, at higher values of $m_{\chi}$ ($m_{\chi}\gtrsim0.5$ GeV), the allowed region for lognormal mass distributions of PBHs with the chosen distribution parameters ($\mu$, $\sigma$) are within the similar allowed region for the case of probability distribution of PBH masses obtained in this work. However, at lower masses of dark matter particles ($m_{\chi}\lesssim 0.5$ GeV) a significant amount of baryon heating is required in order to make the corresponding allowed zone coincide or overlap with the same obtained in Fig.~\ref{fig:beta_mchi}. It is observed from Fig.~\ref{fig:beta_mx}(a) that, for $m_{\chi}\lesssim 0.5$ GeV, the allowed region in Fig.~\ref{fig:beta_mx}(a) for lognormal distribution, only the values of $\mu \lesssim 5\times 10^{14}$ g agrees with the allowed region for the probability distribution of PBH masses as described in this subsection (Eq.~\ref{eq:snd}-\ref{eq:mu_f}). For the case of power law distribution of PBH masses (Fig.~\ref{fig:beta_mx}(b)), the allowed regions for all possible chosen values of distribution parameter $\gamma$ lie within the permissible region corresponds to the probability distribution of PBH masses described in this subsection (Eq.~\ref{eq:snd}-\ref{eq:mu_f}).

\section{Summary and Discussions}\label{summary}
In this work, we use the EDGES 21cm results as a representative of a global 21cm excess trough line to constrain different possible mass distributions of PBHs in the Universe. To this end, in this work we consider types of PBH mass distribution (discussed in the literature). These are lognormal mass distribution and power law mass distribution. Also considered are specific fixed masses (monochromatic) for PBHs with each of these cases for PBH mass, a set of coupled equations are solved where we have included the terms related to baryon and dark matter interaction and the effects of PBH evaporation. These coupled differential equations are then solved simultaneously for the evolution of spin temperature $T_s$, the evolution of baryon temperature $T_b$, DM temperature $T_\chi$ and other quantities required to calculate the brightness temperature of 21cm line, $T_{21}$. Thus the contributions of dark matter - baryon interaction, PBH evaporation are also taken into consideration for the computation of temperature evolutions of $T_{21}$. From these analyses the PBH mass distribution parameters are constrained using the EDGES 21cm results considered in this work. The effects of DM of different masses also play a major role for constraining the parameter space of PBHs. In this regard, the allowed regions of the PBH parameter spaces such as $\beta_{\rm BH} - \mu$, $\beta_{\rm BH} - \sigma$ and $\beta_{\rm BH} - \gamma$ are computed for different fixed values of DM mass. Here, $\beta_{\rm BH}$ represents initial mass fraction of PBHs, $\mu$ and $\sigma$ are mean and variance respectively for lognormal mass distribution and $\gamma$ denotes the power law index for power law mass distribution of PBHs.

With the EDGES like 21cm results considered here, an attempt has been made in this work to construct an expression for the probability distribution of the PBH masses. For this purpose, the range of the $T_{21}$ results around its measured central value (at reionization epoch) is first considered and the weights of each $T_{21}^{z=17.2}$ value (i.e., value of $T_{21}$ at $z \sim 17.2$) within this experimentally obtained range ($-300$mK $\leq T_{21}^{z=17.2}\leq -1000$mK) are estimated and thus a weightage distribution of the $T_{21}^{z=17.2}$ values within this range is constructed. It appears that a skew normal distribution ($G_{21}$) best represents these probabilities. Now for different values of $M_{\rm BH}$ (and a particular value of $\beta_{\rm BH}$), $T_{21}^{z=17.2}$ is computed as described in section~\ref{sect. PBH} - \ref{sect. T21}. The weight factor for the calculated values of $T_{21}^{z=17.2}$ (from the distribution $G_{21}$) are then assigned to the corresponding values of $M_{\rm BH}$. A resulting variation of $T_{21}^{z=17.2}$ with $M_{\rm BH}$ is thus constructed for a particular $\beta_{\rm BH}$ and the process is then repeated to yield different such distributions for different fixed values of $\beta_{\rm BH}$. The results are then fitted to assumed form of a bi-variate distribution of $T_{21}^{z=17.2}(M_{\rm BH},\beta_{\rm BH})$. Thus an analytical map between the spin temperature and the PBH mass distribution is obtained. This map is then utilized to obtain a probability distribution for the PBH masses $M_{\rm BH}$. Such form contains certain parameters, the numerical values of which are found out by suitable $\chi^2$-fitting. 
The bi-variate distribution of $T_{21}^{z=17.2}$ is found to be a combination of an error function and Owen function (Eq.~\ref{eq:owentf}). Using this distribution the allowed region of variation of $\beta_{\rm BH}$ and DM mass $m_{\chi}$ is obtained. We then consider two analytical mass distribution $M_{\rm BH}$ namely lognormal distribution and power law distribution with mass distribution function Eq.~\ref{log} and Eq.~\ref{power law} respectively. Note that, for lognormal distribution, the distribution parameters are the mean $\mu$ and the variant $\sigma$, which ate kept at a certain fixed value. For power law distribution however, there is only one parameter, $\gamma$- the power law index, which is to be kept fixed. The brightness temperatures $T_{21}^{z=17.2}$ are computed using Eqs.~\ref{T_chi} to \ref{T21_1} of sect.~\ref{sect. T21} for different values of $\beta_{\rm BH}$ and $m_{\chi}$ and the upper and lower limits of $T_{21}^{z=17.2}$ are obtained in the $\beta_{\rm BH}-m_{\chi}$ plane for different fixed values of $\mu$ (for the case of lognaomal distribution. Note that kept fixed at $\sigma=0.5$). Similar upper and lower limits are also obtained in the $\beta_{\rm BH}-m_{\chi}$ plane for the case of power law distribution for different chosen values of $\gamma$. These are then compared with the allowed region obtained for the probability distribution of $m_{\rm BH}$ (using $G_{21}$) as proposed in this work.

\vskip 1cm
\noindent{\large \bf Acknowledgements}  

\noindent One of the authors (U.M.) receives her fellowship grant from Council of Scientific \& Industrial Research (CSIR), Government of India as Senior Research Fellow (SRF) with the fellowship Grant No. 09/489(0106)/2017-EMR-I. One of the authors (A.H.) wishes to acknowledge the support received from St. Xavier’s
College, Kolkata and the University Grant Commission (UGC) of the Government of India,
for providing financial support, in the form of UGC-CSIR NET-JRF.

\bibliographystyle{JHEP}
\bibliography{reference}
\end{document}